\documentclass[a4paper,11pt]{article}
\usepackage{jheppub} 
\usepackage{lineno}
\usepackage{braket}
\usepackage{mathtools}
\usepackage[svgnames]{xcolor}
\usepackage{amssymb}
\usepackage{subcaption}
\usepackage{bbm}
\usepackage{hyperref}
\usepackage{graphicx}
\usepackage{float}
\usepackage{tikz}
\usepackage{comment}
\usepackage{enumitem}

\title{\boldmath Comments on Class S(YK)}

\author{Micha Berkooz,}
\author{Trivko Kukolj}
\author{and Josef Seitz}

\affiliation{Department of Particle Physics and Astrophysics, Weizmann Institute of Science, \\ Rehovot 7610001, Israel}

\emailAdd{micha.berkooz@weizmann.ac.il}
\emailAdd{trivko.kukolj@weizmann.ac.il}
\emailAdd{josef-emanuel.seitz@weizmann.ac.il}

\abstract{We present a DSSYK-like interpretation of the Schur half-indices of $\mathcal{N}=2$ $SU(2)$ gauge theories with matter, in the presence of fundamental Wilson lines. The Schur half-indices of these theories can be understood as transition amplitudes in a non-vacuum sector of ordinary DSSYK. In the language of chord diagrams, the half-indices are obtained by summing over diagrams with special segments, which correspond to coherent states of the $q$-oscillator algebra. In addition, we show that the Schur half-index of $SU(2)$ gauge theory with $n_F=4$ fundamental half-hypermultiplets corresponds to the partition function of a particle on the quantum disk.}

\begin{document}

\maketitle
\flushbottom

\section{Introduction}
\label{sec:intro}
Dualities between different theories play an important role in physics. Often, information is easier to extract on one side than the other, and the duality can help uncovering some hidden structure which becomes apparent on the other side. One such example that fits the theme of this work is the AGT correspondence \cite{Alday:2009aq}: the four-point function of Liouville theory has a nontrivial ``outer'' symmetry which arises from $S$-duality of $\mathcal{N}=2$ superconformal QCD. We thus look for dualities in order to elucidate hidden structures in theories. In light of this, Gaiotto and Verlinde have made an interesting observation \cite{Gaiotto:2024kze}. They propose a relation between the two following theories:

On one hand, we have the Sachdev-Ye-Kitaev (SYK) model \cite{Sachdev:2015efa,kitaev2015simple,Maldacena:2016hyu,PhysRevLett.70.3339}, a discordered, chaotic \cite{kitaev2015simple,Maldacena:2016hyu,Cotler:2016fpe,Maldacena:2015waa,you2017sachdev,PhysRevD.94.126010} many-body quantum mechanical model consisting of $N$ Majorana fermions with $p$-local interactions. At low energies, it is known to reduce to (a perturbative description of) JT gravity, which describes the near-horizon dynamics of near-extremal black holes \cite{Sachdev:2015efa,kitaev2015simple,Cotler:2016fpe,Maldacena:2016upp,Saad:2018bqo,Maldacena:2018lmt,Goel:2018ubv,Jensen:2016pah,Sarosi:2017ykf,Polchinski:2016xgd,Blommaert:2024ydx,Blommaert:2024whf}. The model has a double-scaling limit ($N\rightarrow \infty$, $p^2/N$ fixed), known as DSSYK \cite{Cotler:2016fpe,Berkooz:2018qkz,Berkooz:2018jqr}, in which it can be solved exactly using diagrammatic techniques. In particular, the partition function can be solved exactly at all temperatures, as a sum over ``chord diagrams'': the $2n$-th Hamiltonian moment is a sum over chord diagrams (which are essentially Wick contractions connected by lines) with $n$ chords. The fact that DSSYK is both chaotic and solvable makes it an important toy model to study. Next to its gravitational description at low temperatures, there have also been recent attempts at relating its high temperature phase to de Sitter space (see for example \cite{Narovlansky:2023lfz,Narovlansky:2025tpb,Tietto:2025oxn,Verlinde:2024zrh,Lin:2022nss,Susskind:2022bia,Susskind:2022dfz,Susskind:2023hnj,Rahman:2022jsf,Rahman:2024iiu,Rahman:2023pgt,Rahman:2024vyg}). In light of this already very rich story, the proposed duality of \cite{Gaiotto:2024kze} is especially interesting.

On the other hand, we have $4d$ $\mathcal{N}=2$ $SU(2)$ gauge theory. $\mathcal{N}=2$ $SU(2)$ gauge theory is asymptotically free and becomes therefore strongly coupled in the IR. It has been solved by Seiberg and Witten \cite{Seiberg:1994aj,Seiberg:1994rs} (for a review, see \cite{Tachikawa:2013kta}). Their construction can be extended to additional matter and to general, even non-Lagrangian  $\mathcal{N}=2$ superconformal theories, via the class $\mathcal{S}$ construction \cite{Gaiotto:2009we}. Of special interest are superconformal indices, i.e. Witten indices in radial quantization \cite{Rastelli:2014jja}, with some chemical potentials for global symmetries. They are protected from quantum corrections and can thus be computed in the UV, from free fields. For Lagrangian theories, this amounts to a simple counting exercise \cite{Gadde:2020yah}. Tuning the potentials to some special value, one obtains the \text{Schur index}. It counts $\frac{1}{4}$ BPS operators. The Schur index on its own is a very rich object. For example, for a class $\mathcal{S}$ theory on a Riemann surface $\Sigma$, it describes the partition function of $q$-deformed Yang-Mills theory on $\Sigma$ \cite{Gadde:2011ik}. It is also the character of a two-dimensional vertex operator algebra \cite{Beem:2013sza}, conjecturally related to the spectrum of BPS particles on the Coulomb branch \cite{Cordova:2015nma}, and via compactification on $S^1$, one can understand the Schur index as encoding some properties of $SL(2,\mathbb C)$ Chern-Simons theory\footnote{We thank Federico Ambrosino for a discussion on this point.} \cite{Gaiotto:2024osr}, which is related to 3D de Sitter gravity \cite{Witten:1989ip}. The precise quantities that Gaiotto and Verlinde match are:
\begin{equation*}
    \langle Tr H^n \rangle_J \quad\leftrightarrow\quad I_n(q).
\end{equation*}
The left hand side is the $n$-th Hamiltonian moment of DSSYK, while the right hand side describes the Schur half-index (a Schur index on $HS^3\times S^1$ with specific boundary conditions) with $n$ insertions of a fundamental Wilson line. They observe that these quantities match by explicitly computing the two sides. What is the intuition behind that result? At a very basic level, we expect that both DSSYK and the Schur half-index ``know'' of some noncommutative structure. For DSSYK, it is the fact that the transfer matrix description \cite{Berkooz:2018jqr,Berkooz:2018qkz} is that of a $q$-harmonic oscillator. For the Schur index of $SU(2)$, the relation to BPS particles on the Coulomb branch \cite{Cordova:2015nma} gives rise to a noncommutative quantum torus algebra that encodes the Kontsevich-Soybelman wall crossing formulas \cite{Kontsevich:2008fj,Dimofte:2009tm}. It is therefore not entirely surprising that the partition function respectively the index encode some noncommutative structure. The fact that they match exactly is however not obvious, and it is unclear whether this is simply a $q$-incidence or whether there is some deep connection between them. If the latter would be the case, this would be extremely interesting due to the rich connections of both sides to other topics. For example, we could ask what is the gauge theory correspondent to chord diagrams. More particularly, as DSSYK is chaotic, that means that there is some chaotic behaviour encoded in the index. Understanding this chaotic behaviour could contribute to new understanding of SUSY dynamics.

In this paper, our goal will be more modest. We will show that the matching of Gaiotto and Verlinde can be extended to all cases of $\mathcal{N}=2$ $SU(2)$ matter theories that are asymptotically free or conformal. These are $SU(2)$ theories with one to four fundamental hypermultiplets, and the $\mathcal{N}=2^*$ theory (i.e. the theory with an adjoint hypermultiplet). We show that each of these Schur half-indices corresponds to a chord partition function with some extra segments, which can be understood as excited states in the chord Hilbert space. The matching extends to both the infinite temperature partition function (which is now nontrivial, compared to ordinary DSSYK, where it is just one), and to Hamiltonian moments. It is known that the density of each Schur half-index can be understood as measure for some set of orthogonal polynomials, the so-called Askey-Wilson polynomials. These diagonalize a transfer matrix process. We show how to relate these processes to the original chord diagram picture. For $SU(2)$ with two hypermultiplets, we show that the story is especially rich: the Schur half-index also turns out to be the partition function of a particle on the quantum disk of \cite{Almheiri:2024ayc}.\\[4pt]
\noindent We note that a 1D description for the Schur half-index of pure $\mathcal{N}=2$ $SU(N)$ gauge theory has also recently been discussed \cite{Lewis:2025qjq}.\\[4pt]
The paper is structured as follows:
\begin{itemize}
    \item In Section \ref{sec:background}, we review the necessary background. Section \ref{subsec: DSSYK} reviews the DSSYK model and while Section \ref{subsec:schur indices} details the superconformal index with its Schur (half-)index limit. 
    \item In Section \ref{subsec:1D interpret}, we describe the basic idea behind the DSSYK interpretation of the Schur half-index. We develop two different pictures. In the first, we can think of the matter contributions to the Schur half-index as extra segments in corresponding chord diagrams: one segment per fundamental hypermultiplet and two segments per adjoint. We develop this viewpoint in Section \ref{subsubsec: segm picture generals}. In the second picture, discussed in Section \ref{askey wilson picture generals}, we show how to interpret the Schur half-index densities as orthogonal measures for subclasses of Askey-Wilson polynomials. These polynomials give rise to generalized DSSYK-like transfer matrices.
    \item In Section \ref{sec:nF=2}, we flesh out the simplest example: the $SU(2)$ theory with a single hypermultiplet. In Section \ref{subsubsec:segm picture nF=2}, we describe the statistics of the extra segment and discuss that it can be understood as a coherent state of the $q$-harmonic oscillator. In Section \ref{subsec: askey nF=2 picture}, we write down the relevant transfer matrix and derive its relation to the segment picture. The transfer matrix describes only dynamical chords (i.e. not the ones that go into the extra segment). We also discuss two interpretations of these results in Section \ref{sec:nF=2 bulk to boundary map}. 
    \item In Section \ref{sec:nF=4}, we discuss the case of two hypermultiplets. As in the case of $n_F=2$, we first develop the segment picture. Section \ref{subsec:segm picture nF=4} discusses chord diagrams with two special segments. The chord Hilbert space interpretation is an amplitude between two coherent states. In Section \ref{subsec:askey picture nF=4}, we derive the alternative transfer matrix description and prove its relation to the segment picture. In Section \ref{subsec:quantum disk}, we show that there is a third interpretation of the Schur half-index, as the partition function of a particle on the quantum disk. We finish the section by discussing the bulk to boundary map, in Section \ref{subsec: nF=4 discussion}.
    \item In Section \ref{sec: other matter}, we briefly comment on the case of three hypermultiplets, and on the two conformal theories: four fundamental hypermultiplets or one adjoint hypermultiplet. The qualitatively new feature for the adjoint is that there are two segments with the constraint that the number of chords emanating from each of these two segments has to match. This corresponds to a mixed state in the chord Hilbert space, or two entangled wormholes in its gravitational interpretation.
    \item Finally, in Section \ref{outlook}, we conclude with an outlook on possible future directions.
\end{itemize}

\section{Background}\label{sec:background}

In Section \ref{subsec: DSSYK} we briefly review the Sachdev-Ye-Kitaev model and chord diagrams, which serve as the main tool to compute quantities in its double-scaled limit. Section \ref{subsec:schur indices} details the Schur (half-)indices for $SU(2)$ theories of class $\mathcal{S}$. Readers familiar with these topics can skip to Section \ref{subsec:1D interpret} where we discuss the relation between the 0+1D and 4D systems.

\subsection{The double-scaled SYK model}\label{subsec: DSSYK}

\begin{figure}[h]
    \centering
    \includegraphics[width=\linewidth]{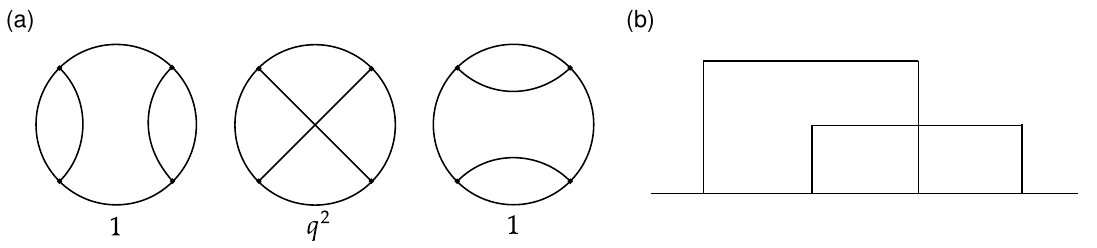}
    \caption{(a) All the chord diagrams contributing to $\langle \text{Tr}(H^4)\rangle_J$. (b) A line representation of the intersecting chord diagram in (a).}
    \label{fig:H4 chord diagram}
\end{figure}

The SYK model \cite{Sarosi:2017ykf,Rosenhaus:2018dtp} is a quantum mechanical model of $N$ Majorana fermions $\psi_i$, $i=1,2,...,N$, satisfying $\{\psi_i,\psi_j\}= 2\delta_{ij}$. Its Hamiltonian consists of all-to-all random $p$-local interactions:
\begin{equation}\label{eqn:H-DSSYK-ferm}
    H = i^{\frac{p}{2}} \sum_{1\leq i_1<...<i_p\leq N} J_{i_1...i_p} \psi_{i_1}...\psi_{i_p}.
\end{equation}
The couplings $J_{i_1...i_p}$ are independent Gaussian random variables, satisfying:
\begin{equation}
    \langle J_{i_1i_2...i_p} \rangle_J = 0, \qquad\quad
    \langle J_{i_1...i_p}J_{j_1...j_p} \rangle_J = \frac{\mathbb{J}^2}{\lambda} \binom{N}{p} ^{-1} \delta^{i_1}_{j_1}...\delta^{i_p}_{j_p}, \text{ with } \lambda = \frac{2p^2}{N}.
\end{equation}
Here, $\langle \cdot \rangle_J$ denotes the ensemble average. We will normalize $\text{Tr}(\mathbbm{1})=1$ and $\mathbb{J}=1$, which corresponds to setting $\langle Tr\; H^2\rangle_J = 1$.\\[5pt]
\noindent This model has been studied extensively in the double-scaled limit ($p\rightarrow \infty,\; N \rightarrow \infty$, fixed $\lambda=\tfrac{2p^2}{N}$), where it can be combinatorially described via \textit{chord diagrams} \cite{Berkooz:2018qkz,Berkooz:2018jqr,Lin:2022rbf,Berkooz:2024lgq,Berkooz:2024ofm,Berkooz:2024evs}. For example, the partition function of the theory at inverse temperature $\beta$ is given by:
\begin{equation}\label{eqn:Z-DSSYK}
    Z(\beta)= \sum_{k=0}^{\infty} \frac{(-2\beta)^{2k}}{(2k)!} \left(\frac{\mathbb{J}^2}{\lambda}\right)^2 \sum_{\substack{\text{chord diagrams} \\ {\text{with $k$ chords}}}} q^{2\cdot \# \text{intersections}}, \qquad \text{ where } q^2 = e^{-\lambda}.
\end{equation}
The inner sum runs over all possible Wick contractions of $2k$ points on a circle. Each contraction corresponds to a single chord diagram, weighed by a factor of $q^2$ for each crossing of two chords\footnote{For easier comparison to the Schur half-index, we use the convention of \cite{Gaiotto:2024kze} where chord intersections are weighed by $q^2$, opposed to \cite{Berkooz:2018jqr,Berkooz:2018qkz}, where the weight is $q$.}. Hamiltonian moments can be obtained by expanding \eqref{eqn:Z-DSSYK} to specific order. E.g. the diagrams contributing to $\langle \text{Tr}\;H^4\rangle_J$ are displayed in Figure \hyperref[fig:H4 chord diagram]{1(a)}.

The sum over chord diagrams grows quickly with $k$, but can be computed efficiently \cite{Berkooz:2018jqr,Berkooz:2018qkz} using (in slight abuse of notation) a transfer matrix $H$. Each chord diagram can be cut open to a line (e.g. Figure \hyperref[fig:H4 chord diagram]{1(b)}), so that it starts and ends with no open chords. The $n$-th Hamiltonian moment corresponds to having $n$ points on a circle, hence by successively applying $H$, we find:
\begin{equation}
    \langle H^{n}\rangle_J = \bra{0} H^{n} \ket{0}.
\end{equation}
The right-hand side represents a vacuum-to-vacuum amplitude in a chord Hilbert space $\mathcal{H}=Span\{\ket{\mathbf{n}}\}$, where $\ket{\mathbf{n}}$ represents a state with $n$ open chords.

\begin{figure}[h]
    \centering
    \includegraphics[width=\linewidth]{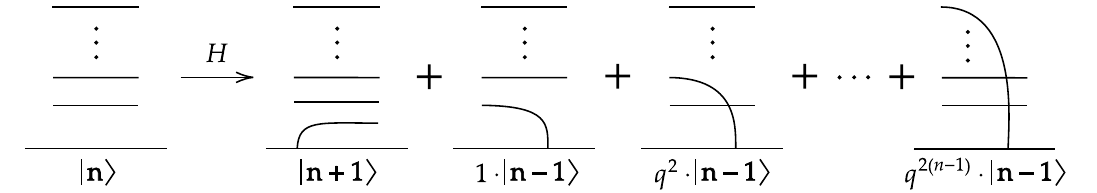}
    \caption{Action of the DSSYK transfer matrix.}
    \label{fig:T}
\end{figure}

A transfer matrix which implements such a shift and correctly accounts for all $q^2$ intersection factors was obtained in \cite{Berkooz:2018qkz}. Intuitively, $H$ implements either opening or closing a single chord. To avoid double-counting any intersections, we adopt a convention where a chord is only opened below all the other open chords, however, any chord can be closed, with different intersection factors (see Figure \hyperref[fig:T]{2}). The transfer matrix $H$ can be written as:
\begin{equation}\label{eqn:T-DSSYK}
    H = a + a^{\dagger}
\end{equation}
where $a^\dagger$ and $a$ are creation and annihilation operators of a q-deformed harmonic oscillator algebra, satisfying $[a,a^\dagger]_{q^2}\equiv aa^\dagger - q^2 a^\dagger a=1$. These operators act on the chord number states as:
\begin{flalign}\label{q Harmonic oscilator relations}
    a\ket{\mathbf{n}} = [n]_{q^2}\ket{\mathbf{n-1}}, \qquad a^\dagger\ket{\mathbf{n}} = \ket{\mathbf{n+1}}, \qquad \braket{\mathbf{n}|\mathbf{m}}= [n]_{q^2}! \delta_{n,m}.
\end{flalign}
where $[n]_{q^2}=\tfrac{1-q^{2n}}{1-q^2}$ (see Appendix \ref{sec: q-Askey scheme}). We call $\{\ket{\mathbf{n}}\}$ the chord basis. We will also utilize the orthonormal basis $\{\ket{n}\}$, where the action of ladder operators and the inner product become:
\begin{flalign}\label{eqn:orthonormal basis}
    a\ket{n} = \sqrt{[n]_{q^2}}\ket{n-1}, \qquad 
    a^\dagger\ket{n} = \sqrt{[n+1]_{q^2}}\ket{n+1}, \qquad 
    \braket{n|m}= \delta_{n,m}.
\end{flalign}
The transfer matrix \eqref{eqn:T-DSSYK} has a continuous spectrum \cite{Berkooz:2018qkz,Berkooz:2018jqr} parametrized by $\theta\in[0,\pi]$:
\begin{equation}
    H\ket{\theta}=\frac{2\cos \theta}{\sqrt{1-q^2}} \ket{\theta}, \qquad
    \ket{\theta} = \sum_{n=0}^\infty \sqrt{(q^2,e^{\pm 2i\theta};q^2)_\infty}\frac{H_n(\cos\theta|q^2)}{\sqrt{2\pi(q^2;q^2)_n}}\ket{n}
\end{equation}
where $H_n(\cos\theta|q^2)$ are continuous q-Hermite polynomials (see Appendix \ref{app:qHermite}). Using these relations, Hamiltonian moments can be recast as integrals over $\theta$:
\begin{equation}\label{eqn:DSSYK-moments}
    \bra{0} H^{2k} \ket{0} = \frac{2}{\pi} \int_0^{\pi} d\theta \sin^2 \theta (q^2,q^2 e^{\pm 2i\theta};q^2)_{\infty} \left(\frac{2\cos\theta}{\sqrt{1-q^2}}\right)^{2k}.
\end{equation}
Note that the zeroth moment is just the partition function at infinite temperature. 

\subsection{\texorpdfstring{$\mathcal{N}=2$}{} theories and Schur (half-)indices}
\label{subsec:schur indices}

For superconformal theories in $d$ dimensions, we can define the superconformal index as the Witten index of the theory in radial quantization on $S^{d-1}\times S^1$. Due to the conformal symmetry, we can think of this object as a sum over local operators $\mathcal{O}$. The $\mathcal{N}=2$ superconformal algebra consists of complex Poincar\'{e} supercharges $\mathcal{Q}_{\alpha}^I, \tilde{\mathcal{Q}}_{I,\dot{\alpha}}$ and conformal supercharges $\mathcal{S}_{\alpha}^I = (\mathcal{Q}_{\alpha}^I)^\dagger, \tilde{\mathcal{S}}_{I,\dot{\alpha}} = (\tilde{\mathcal{Q}}_{I,\dot{\alpha}})^\dagger$, where $I\in\{1,2\}$. As is customary, we will define the index with respect to $\tilde{Q}_{2\dot{-}}$. In four dimensions, the most general expression we can write down depends on three superconformal fugacities $p$, $q$ and $t$ and fugacities $a_i$ for the flavor symmetry $G_F$ \cite{Rastelli:2014jja}:
\begin{equation}\label{superconf index}
    \mathcal{I}(p,q,t,\{a_i\}) \equiv \text{Tr}_{\mathcal{O}}\bigg((-1)^F p^{j_2-j_1-r} q^{j_1+j_2-r} t^{R+r} \prod_{i=1}^{\text{rank} G_F} a_i^{f_i} e^{-\beta \delta_{2,\dot{-}}}\bigg)
\end{equation}
where $2\delta_{2,\dot{-}} = \{ \tilde{\mathcal{Q}}_{2,\dot{-}},(\tilde{\mathcal{Q}}_{2,\dot{-}})^{\dagger} \}$. $j_1,j_2$ are angular momenta of the isometry group $SU(2)_1 \times SU(2)_2$ (with indices $\alpha, \dot{\alpha}$) acting on $S^3$, $R$ the Cartan of $SU(2)_R$ and $r$ the $U(1)_r$ charge.
As it stands, the definition \eqref{superconf index} only makes sense for superconformal theories. It depends explicitly on the $U(1)_r$ charge, which sits inside the same supersymmetric multiplet as the conformal transformations. 

One can take a special limit of the fugacities such that the amount of preserved supersymmetry is enhanced. One choice of parameters for which that happens is $t=q$. The index then becomes independent of $p$ \cite{Rastelli:2014jja,Gadde:2020yah}, so we can set it to zero\footnote{The ordering of limits is subtle: when first setting $p$ to zero and then $t=q$, one gets a different result. For an interesting recent discussion on the consequences of this phenomenon, see \cite{Deb:2025ypl}.}.  In that case, the operators that contribute to the index fulfill two shortening conditions \cite{Rastelli:2014jja}:
\begin{equation}\label{shortening conditions}
    \begin{split}
         & \delta^1_+=\{\mathcal{Q}^1_+, (\mathcal{Q}^1_+)^{\dagger} \} = E+2j_1 - 2R -r = 0 
         \\[5pt]
          & \delta_{2\dot{-}} = \{\tilde{\mathcal{Q}}_{2,\dot{-}}, (\tilde{\mathcal{Q}}_{2,\dot{-}})^{\dagger} \} = E-2j_2 - 2R +r = 0. 
    \end{split}
\end{equation}
The resulting object, known as the \textit{Schur index}, counts $1/4$ BPS operators\footnote{Note that here we have passed from $q\rightarrow q^2$ in comparison to \cite{Rastelli:2014jja}.}:
\begin{equation}\label{Schur index expression}
    \mathcal{I}(q,\{a_i\}) = \text{Tr}_{\mathcal{O}}\left((-1)^F q^{2(R + j_2 -j_1)}\prod_{i=1}^{\text{rank} G_F} a_i^{f_i}\right).
\end{equation}
This expression is independent of $r$ and is thus formally well-defined for non-conformal theories as well. From the superconformal point of view, it can be viewed as counting operators cohomological with respect to a specific combination of supercharges \cite{Rastelli:2014jja}. 

One can also understand the Schur index via a holomorphic-topological twist, on an rigid supergravity background \cite{Festuccia:2011ws,Gaiotto:2024kze,Gaiotto:2024tpl,Kapustin:2006pk,Kapustin:2006hi,Mikhaylov:2017ngi}. In terms of radial quantization on $S^3\times \mathbb{R}$, the ``topological part'' of the twist amounts to restricting to a great circle $S^1 \subset S^3$. Thinking about the Schur index as counting operators on $\mathbb{R}^4 = \mathbb{R}\times \mathbb{R}\times \mathbb C $, the index only captures operators localized at the origin in $\mathbb{C}$, as these are the only operators invariant under a rotation in that plane. In the remaining plane, the ``holomorphic part'' of the twist ensures that we are only considering a chiral half of a theory, with the resulting theory corresponding to a vertex operator algebra \cite{Beem:2013sza}.

As the Schur index is a protected quantity, it can be computed in the free theory limit, by counting gauge-invariant operators (``words'') built out of the fundamental operators (``letters'') present in the $\mathcal{N}=2$ multiplets. The only letters that survive in the Schur limit are vector multiplet gauginos $\lambda$, $\bar{\lambda}$, and the hypermultiplet scalars $Q, \bar{Q}$. Their quantum numbers are displayed in Table \ref{table: letter quantum numbers}.
\begin{table}[H]
\begin{centering}
\begin{tabular}{|c|r|r|r|r|r|c|}
\hline
Letters & $  E$ & $j_1$ & $  j_2$ & $R$ & $r$ & $\mathcal{I}(q)$ \tabularnewline
  \hline
   \hline
$  \lambda^1_{-}$ & $  \frac{3}{2}$ & $  -  \frac{1}{2}$ & $0$ & $  \frac{1}{2}$ & $-  \frac{1}{2}$  &  $-q^2$ \tabularnewline
  \hline
$  \bar{\lambda}_{2\dot{+}}$  & $  \frac{3}{2}$ & $0$ & $  \frac{1}{2}$ & $  \frac{1}{2}$ & $  \frac{1}{2}$ & $-q^2$ \tabularnewline
  \hline
  \hline
$Q,\bar{Q}$ & $1$ & $0$ & $0$ & $  \frac{1}{2}$ & $0$ &  $q$ \tabularnewline
  \hline
    \hline
$  \partial_{-\dot{+}}$ & $1$ & $  -  \frac{1}{2}$ & $  \frac{1}{2}$ & $0$ & $0$  & $q^2$ \tabularnewline
\hline
\end{tabular}
\par  \end{centering}
  \caption{The single particle letters contributing to the Schur index. Note that the conformal dimension $E$ and the $U(1)_r$ charge $r$ are not necessary to define the Schur index, due to the definition \eqref{Schur index expression} and are, in fact, absent in the non-conformal case.}
\label{letters}\label{table: letter quantum numbers}
\end{table}
\noindent It is useful to first build the single letter partition function for each multiplet. This is just the contribution of each individual letter, weighed by the character of the corresponding representation under the gauge group (in our case, $SU(2)$). For matter multiplets, we also include the flavor symmetry characters, for a generic representation $\mathcal{R}$. One finds:
\begin{flalign}
    &f^V(q^2,\theta) = -\frac{2q^2}{1-q^2} \chi_1(\theta), \qquad\qquad\quad\quad\;\; \chi_1(\theta) = e^{2i\theta} + 1 + e^{-2i\theta}.\label{vector multiplet single letter}
    \\&
    f^H_{fund.}(q^2,\theta) = \frac{2q}{1-q^2} \chi_{\frac{1}{2}}(\theta) \chi_\mathcal{R}(\{a_i\}),  \qquad
    \chi_{1/2}(\theta) = e^{i\theta} + e^{-i\theta}. \label{hypermultiplet single letter - fund}
    \\&
    f^H_{adj.}(q^2,\theta) = \frac{2q}{1-q^2} \chi_{1}(\theta) \chi_\mathcal{R}(\{a_i\}).\label{hypermultiplet single letter - adj}
\end{flalign}
The full index is now built by taking the plethystic exponent (see \eqref{eqn:plethystic}) and integrating over the the gauge group with the appropriate Haar measure \cite{Gadde:2020yah}. For example, for $\mathcal{N}=2$ $SU(2)$ theory with $n_F$ fundamental half-hypermultiplets, we have:
\begin{equation}\label{eqn:FullIndex}
    \mathcal{I}_{fund.}(q,\{\gamma_l\};n_F)=8\int_0^{\pi} \frac{d\theta}{4\pi} \sin^2\theta \bigg[(q^2,q^2e^{\pm 2i\theta};q^2)_{\infty} \prod_{l=1}^{n_F/2}\frac{1}{(qe^{\pm i\theta+ i\gamma_l};q^2)_{\infty}}\bigg]^2.
\end{equation}

\paragraph{The Schur half-index.} One can now obtain the Schur half-index \cite{Gaiotto:2024kze} by putting the theory on $\mathbb{R}^+ \times \mathbb{R}\times \mathbb{C}$ \cite{Dimofte:2011py,Cordova:2016uwk}. Once again, the holomorphic-topological twist projects to operators that sit at the origin of $\mathbb{C}$. We then have to choose boundary conditions on the boundary of the half-plane $\mathbb{R}^+ \times \mathbb{R}$. As in \cite{Gaiotto:2024kze}, we choose Neumann boundary conditions for the gauge fields; this implies that the $\mathcal{N}=1$ chiral multiplet inside the $\mathcal{N}=2$ has Dirichlet boundary conditions and does not contribute to the index \cite{Gaiotto:2024kze,Lewis:2025qjq}. Effectively, this means that we drop the factor of two in the single letter partition function \eqref{vector multiplet single letter}. Similarly for hypermultiplets, if we choose Neumann boundary conditions for one of the $\mathcal{N}=1$ chiral multiplets inside the hypermultiplet, the other $\mathcal{N}=1$ multiplet has Dirichlet boundary conditions and does not contribute to the index. Like in the case of the vector multiplet, this removes the factor of two in front of the hypermultiplet single letter partition function \eqref{hypermultiplet single letter - fund}, \eqref{hypermultiplet single letter - adj} \cite{Lewis:2025qjq}. Hence the half-index can essentially be obtained from the full index \eqref{eqn:FullIndex}, by taking the square-root of the integrand.

One can also compute Schur half-indices in the background of fundamental Wilson lines. These can be inserted as local operators along the equator of $HS^3$, while wrapping the $S^1$. From the point of view of the theory of the half-plane, the Wilson lines are topological and stretch from the origin of $\mathbb{R}^+ \times \mathbb{R}$ radially outwards. Each such insertion contributes a factor of $(\tfrac{\chi_{1/2}(\theta)}{\sqrt{1-q^2}})$ to the half-index, in analogy to the $n_F=0$ case \cite{Gaiotto:2024kze}.\\

\noindent For $SU(2)$, there are only five Lagrangian theories which are asymptotically free or conformal: the theory with $n_F=2,4,6,8$ fundamental half-hypermultiplets, and the theory with an adjoint hypermultiplet (i.e. the $\mathcal{N}=2^*$ theory). There are no theories with an odd number of half-hypermultiplets due to the Witten anomaly \cite{Witten:1982fp}. For $n_F$ hypermultiplets in the fundamental representation, we have an $SO(n_F)$ flavor symmetry. The boundary condition we impose breaks that flavor symmetry to $SO(\lfloor \frac{n_F}{2} \rfloor)$. Hence, we write the Schur half-index with $n_F$ fundamental half-hypermultiplets in the presence of $n$ fundamental Wilson lines as:
\begin{equation}\label{eqn:halfIndex-fund}
    I_{n}(q,\{\gamma_l\},n_F)=8\int_0^{\pi} \frac{d\theta}{4\pi} \sin^2 \theta (q^2,q^2e^{\pm 2i\theta};q^2)_{\infty} \prod_{l=1}^{n_F/2}\frac{1}{(qe^{\pm i\theta+ i\gamma_l};q^2)_{\infty}} \bigg(\frac{2\cos\theta}{\sqrt{1-q^2}}\bigg)^n,
\end{equation}
where chemical potentials $\gamma_l$ are turned on according to the $SO(\lfloor \frac{n_F}{2}\rfloor)$ symmetry. Similarly, for the theory with a single adjoint hypermultiplet, the Schur half-index with $n$ fundamental Wilson lines reads:
\begin{equation}\label{eqn:halfIndex-adj}
    I^{adj}_{n}(q)= 8\int_0^{\pi} \frac{d\theta}{4\pi} \sin^2 \theta (q^2,q^2e^{\pm 2i\theta};q^2)_{\infty}\frac{1}{(q,q e^{\pm 2i\theta};q^2)_{\infty}} \bigg(\frac{2\cos\theta}{\sqrt{1-q^2}}\bigg)^n.
\end{equation}

\noindent The physics of the Schur (half-)index with Wilson lines inserted is just that of the Kondo chain \cite{Affleck:1995ge,Tsvelik:1996zj,Harrison:2011fs}. In discussions of the Kondo impurity and Kondo chains in the context of condensed matter physics, the issue is that the ambient fermions screen the spin of the impurity. However, since we inserted an array of spin 1/2 impurities (through $\chi_{1/2}(\theta)$), the adjoint fermions cannot screen each impurity separately. This is clear when we write the measure of the Schur half-index in terms of characters:
\begin{equation}
    (q^2;q^2)_{\infty}(q^2e^{2i\theta};q^2)_{\infty}(q^2e^{-2i\theta};q^2)_{\infty} = \sum_{n=0}^{\infty} (-1)^n q^{n(n+1)}\chi_n(\theta).
\end{equation}
The factor of $(-1)^n q^{n(n+1)}$ is just the statement that a spin $n$ representation can be screened by $n$ fermions in the adjoint. Since they are fermions, they need a larger and larger Fermi sea. The number of fermions gives a factor of $q^{2n}$ and the larger Fermi sea gives a factor of $q^{2(0+1+2+..(n-1))}$. The $(-1)^n$ is just $(-1)^F$ of the Fermi surface.

\section{DSSYK interpretation of Schur half-indices}
\label{subsec:1D interpret}

As observed in \cite{Gaiotto:2024kze}, the Schur half-index of pure $\mathcal{N}=2$ $SU(2)$ gauge theory matches the partition function of DSSYK at infinite temperature, with insertions of fundamental Wilson lines corresponding to DSSYK Hamiltonian moments. This is just the equality of \eqref{eqn:DSSYK-moments} and \eqref{eqn:halfIndex-fund} for $n_F=0$. It is then natural to ask what interpretation in terms of standard chord rules interpretation we can give to $SU(2)$ theories with matter. By {\it standard chord rules} we mean that we propagate the system with the ordinary DSSYK transfer matrix. 

The reason to think that this is possible is the following. Recall that matter contributes bosonic particles to the index and the screening picture mentioned above suggests that they can participate in screening/dressing. 
In this case, there is no exclusion principle at work, which suggests that the bosons can dress each Kondo impurity independently. Loosely speaking, we expect to have the dynamics of standard chord rules (which corresponds to the impurity and Fermi surface screening) dressed with a diagonal term in the transfer matrix which has to do with the dressing of each impurity by the bosons. 

We would like to make this more precise. Even irrespective of the screening/dressing arguments, it will give us of a way of understanding the matter Schur half-indices within the DSSYK framework. The main point is that we can work in two different pictures:
\begin{itemize}
    \item In the segment picture, we write the expression for the Schur half-index as an initial/final state in DSSYK, and use the standard DSSYK transfer matrix. Loosely, the DSSYK evolution takes care of the dressing by the Fermi sea and the initial/final states take care of the impurity dressing by the bosons.
    \item In the Askey-Wilson picture, we use a more complicated transfer matrix obtained from the recursion relation of orthogonal polynomials in the q-Askey scheme. The orthogonality measure of these polynomials is precisely the density of the Schur half-index with matter. One can relate the transfer matrices of the two pictures to another.
\end{itemize}

\subsection{The segment picture}
\label{subsubsec: segm picture generals}

In the segment picture\footnote{We focus on theories with fundamental matter, as the $\mathcal{N}=2$ theory with a single adjoint hypermultiplet will be treated separately in Section \ref{subsec: adjoint matter}.}, we think of \eqref{eqn:halfIndex-fund} as a sum over standard chord diagrams with $n$ DSSYK Hamiltonian chords and some extra segments that have chords emanating from them. We use the generating function \eqref{q Hermite generating functional} to rewrite the Schur half-index as an integral over products of q-Hermite polynomials:
\begin{equation}\label{q Hermite generating functional}
    \frac{1}{(q e^{i\gamma_l} e^{\pm i \theta};q^2)_{\infty}} = \sum_{k=0}^{\infty} \frac{(qe^{i\gamma_l})^k}{(q^2;q^2)_k} H_k(\cos\theta|q^2).
\end{equation}
If we have $n_F$ half-hypermultiplets we will have the product of $n_F/2$ such sums.  To reach a chord diagram interpretation of the half-index, we need to consider the integral with the measure that corresponds to $n_F=0$, but with $n_F/2$ q-Hermite polynomials inserted. We can then use the following identity \cite{Ismail2005,8cc0700ea80a492fa61a34ea71833640}:
\begin{equation}\label{Hermite chord identity}
    \int_0^{\pi} \frac{d\theta}{2\pi} (q^2,e^{\pm 2i \theta};q^2)_{\infty} \prod_{k=1}^m \frac{H_{n_k}(\cos\theta|q^2)}{\sqrt{1-q^2}^{n_k}}= \sum_{n_{ij}} \prod_{k=1}^m \binom{n_k}{n_{1k},\dots, n_{nk}}_{q^2} \prod_{1\leq i<j \leq m} [n_{ij}]_{q^2}! q^{2B}.
\end{equation}
The sum on the right hand side runs over $m\times m$ symmetric matrices satisfying $n_{ii}=0$, whose $i$-th column sums to $n_i$. $B$ is defined as:
\begin{equation}
    B = \sum_{1\leq i<j<k<l\leq m} n_{ik} n_{jl}.
\end{equation}
The meaning of this formula is the following. We divide the circle into $m$ segments, each having $n_i$ points. \eqref{Hermite chord identity} counts the number of perfect matchings (weighted by factors of $q^2$ for each intersection) of this configuration - i.e., pairing by chords of points in the different segments such that chords are not allowed to start and end in the same segment. The $n_{ij}$ as the number of chords going from the $i$-th to the $j$-th segment. For a more detailed discussion of the combinatorics associated with the right hand side see also \cite{Berkooz:2024ofm,Berkooz:2024evs}.

\noindent Consider now: 
\begin{equation}\label{Hermite chord with defects}
    \int_0^{\pi} \frac{d\theta}{2\pi} (q^2,e^{\pm 2i \theta};q^2)_{\infty}\; H_{1}^n(\cos\theta|q^2)\; \prod_{k=1}^{\frac{n_F}{2}} \frac{H_{n_k}(\cos\theta|q^2)}{\sqrt{1-q^2}^{n_k}}.
\end{equation}
As $H_1(\cos\theta|q^2)=2\cos\theta$, we have $n$ insertions of points on which a chord can start and end without restrictions, and then a set of $\frac{n_F}{2}$ segments with chords that don't start and end in the same segment. We will refer to the $H_1^n(\cos\theta|q^2)$ as the \textit{dynamical region} and refer to the $\frac{n_F}{2}$ special segments as {\it reservoir segments}. For concrete examples, see Figure \hyperref[fig:nF=2 segment]{3} for $n_F=2$, and Figure \hyperref[fig:nF=4 transfer matrix]{4} for $n_F=4$.

This segment picture also has a Hilbert space description, as a non-vacuum process of the DSSYK transfer matrix. Ordinary DSSYK moments are described by vacuum to vacuum amplitudes in the chord Hilbert space. We can then try to think of the Schur half-indices for theories with matter as amplitudes:
\begin{equation}
    I_{n}(q,\{\gamma_l\},n_F) \cong \bra{\Psi} H^n \ket{\Phi},
\end{equation}
where the right hand side is computed in the DSSYK chord Hilbert space. We can be slightly more explicit, by using the fact that the (normalized) basis vectors in DSSYK are given as \cite{Berkooz:2018qkz}:
\begin{equation}
    \braket{k|\theta} = \sqrt{\frac{(q^2;q^2)_{\infty} (e^{\pm 2i\theta};q^2)_{\infty}}{2\pi}} H_k(\cos\theta|q^2).
\end{equation}
Then:
\begin{equation}\label{general state transfer matrix}
\scalebox{0.9}{$
    \begin{aligned}
        \bra{\Psi} H^n \ket{\Phi} & = \int_0^{\pi}d\theta \braket{\Psi|\theta} \left(\frac{2\cos \theta}{\sqrt{1-q^2}}\right)^n \braket{\theta|\Phi}\\ & = \int_0^{\pi} d\theta \frac{(q^2;q^2)_{\infty} (e^{\pm 2i\theta};q^2)_{\infty}}{2\pi} \left(\frac{2\cos \theta}{\sqrt{1-q^2}}\right)^n \sum_{k=0}^{\infty} \Phi_k^* H_k(\cos \theta|q^2)\sum_{k'=0}^{\infty} \Psi_{k'} H_{k'}(\cos \theta|q^2).
    \end{aligned}
$}
\end{equation}
The coefficients $\Phi_k$, $\Psi_k$ can now be engineered to give the matter contribution to the Schur half-index density. Intuitively, one can think of the $H_n$'s as $n$-chord operators and the states $\ket{\mathbf{n}}$ \eqref{q Harmonic oscilator relations} as the corresponding states. In that sense, computing the sum over chord diagrams amounts to computing a correlation function of operators. 

\subsection{The Askey-Wilson picture}
\label{askey wilson picture generals}

A second point of view is to interpret the Schur half-index as corresponding to a generalized SYK-like system. We write \eqref{eqn:halfIndex-fund} with $n$ fundamental Wilson lines inserted, as some $\bra{0}T_{n_F}^n \ket{0}_{n_F}$, with $T_{n_F}$ being a more complicated transfer matrix in the chord Hilbert space. In \eqref{eqn:halfIndex-fund}, we constrain ourselves to fugacities $e^{i \gamma_l}$ that are either real or come in conjugate pairs, so that the half-index is a real object. This restriction corresponds exactly to the $SO(\lfloor \frac{n_F}{2}\rfloor )$ flavor subgroup discussed in Section \ref{subsec:schur indices}. The Schur half-index density is exactly the measure for which specific classes of Askey-Wilson polynomials are orthogonal (see Appendix \ref{sec: q-Askey scheme}). These polynomials (let's call them $p_n(\theta)$) schematically fulfill a recursion relation of the type:
\begin{equation}
    \frac{2\cos \theta}{\sqrt{1-q^2}} p_n(\theta) = a_n p_{n+1}(\theta) +b_n p_n(\theta) + c_n p_{n-1}(\theta). 
\end{equation}
We can then define a transfer matrix on the chord Hilbert space $\mathcal{H}= Span\{\ket{n}\}$ as:
\begin{equation}
    T_{n_F}\ket{n} = \tilde{a}_n \ket{n+1} + \tilde{b}_n \ket{n} + \tilde{c}_n \ket{n-1},
\end{equation}
(with $\tilde{a}_n, \tilde{b}_n, \tilde{c}_n$ related to $a_n,b_n,c_n$) and its eigenstates $\ket{\theta}_{n_F} = \sum_{n=0}^\infty p_n(\theta) \ket{n}$ with eigenvalues $\tfrac{2\cos\theta}{\sqrt{1-q^2}}$.  $\rho_{\text{half-index}} =|\braket{\theta|0}_{n_F}|^2$ is just the Schur half-index density. So after appropriately normalizing the vectors, we find the transfer matrix moments:
\begin{equation}\label{eqn:AWpicture-T}
    \begin{split}
       \bra{0} T_{n_F}^n \ket{0}_{n_F} & =  \int_0^{\pi} d\theta \bra{0}T_{n_F}^n \ket{\theta}_{n_F} \braket{\theta|0}_{n_F} = \int_0^{\pi} d\theta \braket{0|\theta}_{n_F} \left(\frac{2\cos \theta}{\sqrt{1-q^2}}\right)^n \braket{\theta|0}_{n_F}\\  
     &  =\int_0^\pi d\theta \rho_{\text{half-index}} \left(\frac{2\cos \theta}{\sqrt{1-q^2}}\right)^n. 
    \end{split}
\end{equation}
For $n_F=0$, $p_n(\theta)$ are the well-known q-Hermite polynomials $H_n(\cos\theta|q^2)$.

\section{\texorpdfstring{$n_F=2$}{}}\label{sec:nF=2}
For $n_F=2$, we have one full matter hypermultiplet. The Schur half-index with $n$ insertions of fundamental Wilson lines is given by:
\begin{equation}\label{eqn:nf=2 half-index}
    I_{n}(q,n_F=2) = \int_0^{\pi} \frac{d\theta}{2\pi} (q^2,e^{\pm 2i\theta};q^2)_{\infty} \frac{1}{(qe^{\pm i \theta};q^2)} \left( \frac{2\cos\theta}{\sqrt{1-q^2}}\right)^n.
\end{equation}
Following Section \ref{subsec:schur indices}, there is no flavor symmetry due to the half-index projection. We discuss the chord interpretation of \eqref{eqn:nf=2 half-index}, along the general discussion of Section \ref{subsec:1D interpret}.

\subsection{The segment picture}
\label{subsubsec:segm picture nF=2}

By \eqref{q Hermite generating functional}, we can write the half-index as: 
\begin{equation}\label{segment picture Hamiltonian moments}
    I_{n}(q,n_F=2) = \sum_{k=0}^{\infty} \left(\frac{q}{\sqrt{1-q^2}}\right)^k \frac{1}{[k]_q!} \left\langle H_k(x|q^2) \left( \frac{2\cos \theta}{\sqrt{1-q^2}}\right)^n \right\rangle_{n_F=0}.
\end{equation}
The right hand side of \eqref{segment picture Hamiltonian moments} can be rewritten as a sum over chord diagrams (see \eqref{Hermite chord identity}). Compared to $n_F=0$, each chord diagram has an additional segment with $k$ chords sprouting to other segments. This is the reservoir, and the remaining segments form the dynamical region. Because of the factor $\frac{1}{[k]_{q^2}!}$ in \eqref{segment picture Hamiltonian moments}, we do not count the $k!$ permutations of the reservoir chords, but sum only over the diagrams where the reservoir chords do not intersect themselves.
\begin{figure}[h]
    \centering
    \includegraphics[width=\linewidth]{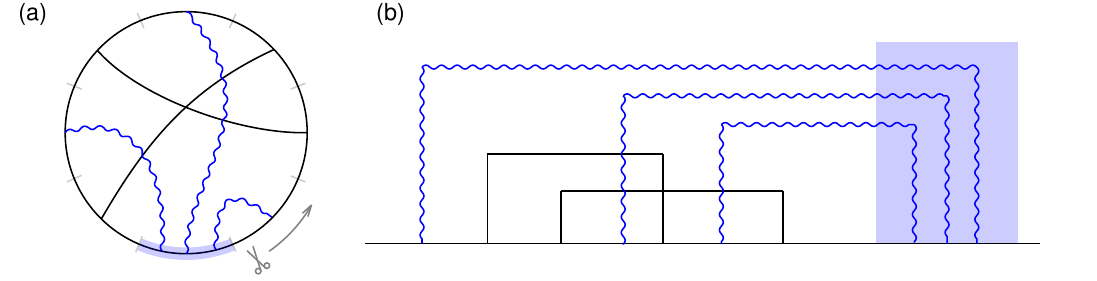}
    \caption{A chord diagram contributing to $\bra{0}T_{n_F=2}^{ 7}\ket{0}$. The reservoir is shaded in blue. Dynamical chords are drawn in straight lines, and reservoir chords are drawn in blue squiggly lines. (a) The reservoir is a segment from which an arbitrary number of chords emanate into the dynamical region. (b) The diagram corresponds to a transfer matrix process, where generating a reservoir chord corresponds to the diagonal term in $T_{n_F=2}$.}
    \label{fig:nF=2 segment}
\end{figure}
By cutting open diagrams in Figure \hyperref[fig:nF=2 segment]{3(a)} to the right of the reservoir, we see that the chord Hilbert space perspective is to start with zero open chords (i.e. $\ket{0}$ as initial state), propagate with $H_1^n(\cos\theta|q^2)$ along the dynamical region, and then end with a state $\ket{\Psi}$ that has some number of open chords $\ket{\mathbf{k}}$. The distribution over $k$ is given by the coefficient of the sum over $k$ in \eqref{segment picture Hamiltonian moments}:
\begin{equation}
    I_{n}(q,n_F=2) = \sum_{k=0}^{\infty} \left(\frac{q}{\sqrt{1-q^2}}\right)^k \frac{1}{[k]_{q^2}!} \bra{\mathbf{k}} H^n \ket{0} \equiv \bra{\Psi_0} H^n \ket{0}.
\end{equation}
Here we have defined the \textit{1-reservoir state}:
\begin{equation}\label{eqn:1-reservoir}
    \ket{\Psi_{\gamma}} = \sum_{k=0}^{\infty} \left(\frac{qe^{i\gamma}}{\sqrt{1-q^2}}\right)^k \frac{1}{[k]_{q^2}!} \ket{\mathbf{k}}. 
\end{equation}
We allow for a phase $\gamma$ since such a state will appear for larger $n_F$, where $\gamma$ is a chemical potential for a flavor symmetry. 
$\ket{\Psi_{\gamma}}$ is a coherent state of the $q$-harmonic oscillator:
\begin{equation}
    a\ket{\Psi_{\gamma}} = \sum_{k=0}^{\infty} \left(\frac{q e^{i\gamma}}{\sqrt{1-q^2}}\right)^k \frac{1}{[k]_{q^2}!} a \ket{\mathbf{k}} = \frac{qe^{i\gamma}}{\sqrt{1-q^2}} \ket{\Psi_{\gamma}},
\end{equation}
where we've used \eqref{q Harmonic oscilator relations}. It has an $a^{\dagger} a$-expectation value:
\begin{equation}\label{adaggera exp value}
    \frac{\bra{\Psi_{\gamma}} a^{\dagger} a \ket{\Psi_{\gamma}}}{\braket{\Psi_{\gamma}| \Psi_{\gamma}}} = \frac{q^2}{1-q^2},
\end{equation}
which diverges as $1/\lambda$ when $q^2 = e^{-\lambda} \rightarrow 1$. \eqref{adaggera exp value} has an interesting gravitational interpretation that we will explore momentarily. The occupation number is independent of $\gamma$ and only enters in observables like $a+a^{\dagger}$. The fact that $\ket{\Psi_{\gamma}}$ is a coherent state is interesting, since it allows us to borrow some intuition about these being solitonic (or semiclassical) states in the model. 

Each time a chord goes to the reservoir, it just returns a factor of $q/\sqrt{1-q^2}$, without influencing later chords that also go to the reservoir. This suggests that we should be able to keep track only of the dynamics of chords that do not enter the reservoir, and account for the reservoir by some correction to the transfer matrix. It turns out that the Askey-Wilson picture precisely realizes that intuition.

\subsection{The Askey-Wilson picture}
\label{subsec: askey nF=2 picture}

As explained in Section \ref{askey wilson picture generals}, we interpret the Schur index density as measure for some orthogonal polynomials. In the case of $n_F=2$, the relevant orthogonal polynomials are generalized $q$-Hermite polynomials $\phi(\cos\theta,t_1|q^2)$. We have summarized some of their properties in Appendix \ref{generalized q Hermite}. They fulfill the recursion relation:
\begin{equation}\label{gen q Hermite recursion}
    \frac{2\cos \theta}{\sqrt{1-q^2}} \phi_n = \sqrt{[n+1]_{q^2}} \phi_{n+1} + \frac{t_1 q^{2n}}{\sqrt{1-q^2}} \phi_n + \sqrt{[n]_{q^2}} \phi_{n-1},
\end{equation}
(here, we set $t_1=q$).
In this convention, $\phi_n = \braket{n|\theta}_{n_F=2}$ are normalized both with respect to the summation over $n$ as well as integration over $\theta$. Let us now define the transfer matrix 
\begin{equation}\label{nf=2 transfer matrix}
    \begin{split}
        T_{n_F=2}\ket{n} & = \sqrt{[n+1]_{q^2}!}\ket{n+1} + \sqrt{[n]_{q^2}!} \ket{n-1} + \frac{t_1}{\sqrt{1-q^2}} q^{2n} \ket{n} \\
        & = (a + a^{\dagger} + \frac{t_1}{\sqrt{1-q^2}}q^{2n}) \ket{n}.
    \end{split}
\end{equation}
We can understand the recursion relation \eqref{gen q Hermite recursion} as eigenvector equation for $\ket{\theta}_{n_F=2}$ with eigenvalue $\frac{2\cos \theta}{\sqrt{1-q^2}}$. The $\phi_n$ are its components in the $\ket{n}$-basis, $\braket{n|\theta}$. Then we can compute the vacuum to vacuum transfer matrix process:
\begin{equation}
    \begin{split}
        \bra{0} T_{n_F=2}^n \ket{0} & = \int_{0}^{\pi} d\theta |\phi_0(\theta)|^2 \left(\frac{2\cos \theta}{\sqrt{1-q^2}} \right)^n = \frac{1}{2\pi}\int_0^{\pi} d\theta \frac{(q^2,e^{\pm 2i\theta};q^2)_{\infty}}{(q e^{\pm i \theta};q^2)_{\infty}} \left(\frac{2\cos \theta}{\sqrt{1-q^2}} \right)^n \\ & = I_{n}(q,n_F=2).
    \end{split}
\end{equation}
In the first equality, we have inserted a decomposition of unity into the $\ket{\theta}_{n_F=2}$ basis, while in the second equality, we have used the explicit form of the wavefunctions from Appendix \ref{generalized q Hermite}. The transfer matrix \eqref{nf=2 transfer matrix} therefore describes the $n_F=2$ Schur half-index. Its form can be understood as follows:

Consider a chord diagram with some number of reservoir chords. We cut the diagram open such that the reservoir is to the right (see Figure \ref{fig:nF=2 segment}). At each step in the transfer matrix process, we then have three choices:
\begin{enumerate}
    \item create a dynamical chord 
    \item annihilate a dynamical chord
    \item create a reservoir chord. 
\end{enumerate} 
We denote the number of dynamical chords by $n$ and the corresponding Hilbert space vector as $\ket{\mathbf{n}}$. Creating or annihilating a dynamical chord happens in the same way as in DSSYK and gives an $a + a^{\dagger}$ piece for the transfer matrix. Creating a reservoir chord, we pick up a factor of $t_1/\sqrt{1-q^2}$, where $t_1=q$ (either reading it off from \eqref{segment picture Hamiltonian moments} or by recalling that the corresponding state is coherent), as well as a factor of $q^{2n}$ since the reservoir chord crosses all currently open dynamical chords. We therefore compute the vacuum to vacuum amplitude of the transfer matrix:
\begin{equation}
    T_{n_F=2} = a + a^{\dagger} + \frac{t_1}{\sqrt{1-q^2}}q^{2n}, 
\end{equation}
which is exactly the transfer matrix we found from the Askey-Wilson polynomials. The Askey-Wilson Hilbert space is thus the space of dynamical chords.\\[5pt]

\subsection{DSSYK and spacetime interpretations}\label{sec:nF=2 bulk to boundary map}

We briefly comment about possible interpretations of the combinatorical model in this section, beyond the chord Hilbert space.

\paragraph{A non-unitary deformation of DSSYK}
A simple microscopic model of the $n_F=2$ half-index is an SYK model coupled to a bath. More precisely, we consider a Hamiltonian of the type:
\begin{flalign}
    H= H_{DSSYK} + \mathcal{O}, \qquad\qquad \mathcal{O} = \sum_{1\leq i_1...\leq i_p \leq N} \tilde{J}_{i_1.. i_p}\psi_{i_1}...\psi_{i_p},
\end{flalign}
where $\tilde{J}_{i_1,...,i_p}\in \mathbb{C}$. Here $H_{DSSYK}$ denotes the ordinary SYK Hamiltonian \eqref{eqn:H-DSSYK-ferm} in the double-scaled limit. We choose the gaussian couplings $J$ and $\tilde{J}$ so that $\langle \tilde{J}_{i_1...i_p} \rangle_{J,\tilde{J}} = \langle \tilde{J}_{i_1...i_p}\tilde{J}_{j_1...j_p} \rangle_{J,\tilde{J}} =0$, but $\langle \tilde{J}_{i_1...i_p}\tilde{J}^*_{j_1...j_p} \rangle_{J,\tilde{J}} = \langle J_{i_1...i_p} J_{j_1...j_p} \rangle_{J,\tilde{J}}$, with $\langle\cdot\rangle_{J,\tilde{J}}$ denoting the ensemble average. Then:
\begin{equation}
    \left\langle H^n  \text{exp}_{q^2} \left(\tfrac{t_1}{\sqrt{1-q^2}}\mathcal{O}^{\dagger}\right) \right\rangle = I_n(q;n_F=2),
\end{equation}
where $\exp_{q^2}$ is the $q$-deformed exponential \eqref{eqn:qexp}. The dynamical chords are given by contractions of $H_{DSSYK}$, while the reservoir chords are contractions of $\mathcal{O}$ with $\mathcal{O}^{\dagger}$. The number of reservoir chords then mirrors a dissipation into the bath. 
 
\paragraph{EOW brane in DSSYK} One can understand the reservoir segment \eqref{eqn:1-reservoir} as an external source (the blue shaded region in Figure \ref{fig:nF=2 segment}) which couples to ordinary DSSYK. As a coherent state of the q-oscillator algebra, it injects semiclassical excitations (reservoir chords) into the system. In fact, the reservoir for any $t_1$ has a precise gravitational interpretation as an end of the world (EOW) brane in DSSYK \cite{Okuyama:2023byh}. 
The brane tension $\mu$ is related to $t_1$ as\footnote{In our conventions, $q^2=e^{-\lambda}$, whereas \cite{Okuyama:2023byh} uses $q=e^{-\lambda}$.} $t_1 = q^{2\mu +1}$. The Schur half-index is reproduced for $t_1=q$, which is the limit where the EOW brane becomes tensionless. Its length can be computed in the triple-scaling limit, where $n$ localizes to its semi-classical value: 
\begin{flalign}\label{eqn:tripleScaling}
    t_1=q^{2\mu+1} \quad 
    q^{2n} = \lambda e^{-L}, \quad 
    \theta = \lambda k, \quad 
    \lambda \to 0; \qquad 
    \mu, L, k = \text{fixed}.
\end{flalign}
We find the EOW brane length from the coherent state description:
\begin{equation}\label{EOW brane length}
    \langle e^{-L} \rangle = 2\mu + 1 + \mathcal{O}(\lambda),
\end{equation}
which is finite in the triple scaling limit. In the limit \eqref{eqn:tripleScaling}, the recursion relation \eqref{nf=2 transfer matrix} takes the form of a particle in a Morse potential \cite{Okuyama:2023byh}, which is precisely the Schr\"{o}dinger equation for JT gravity with an EOW brane. The reservoir chords can be understood as generating the EOW brane. States in the Askey-Wilson picture are a microscopic description of the JT Hilbert space in the presence of the brane. We have the identification:
\begin{equation}\label{nF=2 bulk to boundary map}
    \bra{0} T^n \ket{0} = \bra{\Psi_0} H^n \ket{0},
\end{equation}
where the left hand side has a boundary description, as a particle in a Morse potential, while the right hand side can be understood \cite{Lin:2022rbf} as computing a bulk gravitational path integral between a wormhole of length \eqref{EOW brane length} and one of length zero. For $\mu =0$, both descriptions are captured by the half-index. 

\section{\texorpdfstring{$n_F=4$}{}}
\label{sec:nF=4}
The main focus of the paper is the case of four half-hypermultiplets. For $n_F=4$, the half-index has an $SO(2)$ flavor symmetry, parametrized by an angle $\gamma$. The Schur half-index with $n$ fundamental Wilson line insertions is therefore: 
\begin{equation}\label{nF=4 Schur index}
    I_{n}(q,\gamma,n_F=4) = \int_0^{\pi} \frac{d\theta}{2\pi} (q^2,e^{\pm 2i\theta};q^2)_{\infty} \frac{1}{(qe^{i\gamma}e^{\pm i\theta};q^2)} \frac{1}{(qe^{-i\gamma}e^{\pm i\theta};q^2)} \left( \frac{2\cos\theta}{\sqrt{1-q^2}}\right)^n.
\end{equation}
We now describe the segment and Askey-Wilson pictures, as in the case of $n_F=2$, and also describe the quantum disk \cite{Almheiri:2024ayc} interpretation special to $n_F=4$. 

\subsection{The segment picture}
\label{subsec:segm picture nF=4}

As before, we can expand the denominators of the half-index density in $q$-Hermite polynomials and write the $n_F=4$ index as:
\begin{flalign}\label{segment picture Hamiltonian moments nF=4}
\begin{aligned}
    I_{n}(q,\gamma,n_F=4) = & \sum_{k,k'=0}^{\infty} \left(\frac{q e^{i\gamma}}{\sqrt{1-q^2}}\right)^k \left(\frac{q e^{-i\gamma}}{\sqrt{1-q^2}}\right)^{k'} \frac{1}{[k]_q! [k']_q!} \\& 
    \qquad\quad\times  
    \left\langle H_k(\cos\theta|q^2) H_{k'}(\cos\theta|q^2) \left( \frac{2\cos \theta}{\sqrt{1-q^2}}\right)^n \right\rangle_{n_F=0}.
\end{aligned}
\end{flalign}
Using the formula \eqref{Hermite chord identity}, it is clear that we now have two reservoirs. From this rewriting, the factors that correspond to each reservoir chord are now twisted by the $SO(2)$ chemical potential as $t_{1,2}=qe^{\pm i\gamma}$.

\begin{figure}[h]
    \centering
    \includegraphics[width=\linewidth]{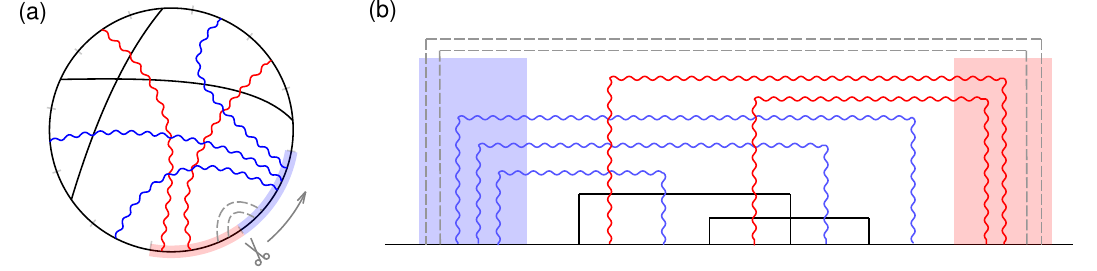}
    \caption{A chord diagram contributing to $\bra{0}T_{n_F=4}^9\ket{0}$. $t_1$-reservoir chords are drawn in blue, $t_2$-reservoir chords are drawn in red. Dynamical chords are drawn in straight lines. The dashed chords stretch between the two reservoirs, without any intersections, hence they only contribute an overall factor.}
    \label{fig:nF=4 transfer matrix}
\end{figure}

The chord Hilbert space description is straightforward to acquire. We cut open the chord diagrams between the two special segments (see Figure \ref{fig:nF=4 transfer matrix}). There are now three types of chords: chords that go from reservoir one to reservoir two, chords that only go to reservoir one or reservoir two, and chords that do not got into the reservoirs at all. We exclude the reservoir-to-reservoir chords for a moment. The chord Hilbert space amplitude is then an amplitude from an excited state to another excited state (contrary to $n_F=2$, where we propagated from the vacuum to an excited state). We can simply apply the calculation of the $n_F=2$ case twice (on both ends of the propagation, with general $\gamma$) and find that: 
\begin{equation}\label{nF=4 state eq}
    I_{n}(q,\gamma,n_F=4) = \bra{\Psi_{\gamma^*}} H^n \ket{\Psi_{\gamma}}.
\end{equation}
We have included a complex conjugate of $\gamma$ for the case that $\gamma$ is not real. For the $n_F=4$ case, we thus propagate from coherent to coherent state. In the way we have defined it, the norm of $\ket{\Psi_{\gamma}}$ is not one. There is an overall factor $\braket{\Psi_{\gamma^*}|\Psi_{\gamma}}$ in \eqref{nF=4 state eq} that we have to explain. 
It can exactly be accounted for by the chords that span from reservoir to reservoir: consider a chord diagram contributing to $I_{n}(q,\gamma,n_F=4)$ with $k_1$ $t_1$-chords, $k_2$ $t_2$-chords and zero chords going from reservoir one to reservoir two. By \eqref{Hermite chord identity}, such a diagram contributes to $I_{n}(q,\gamma,n_F=4)$ as: 
\begin{equation}
    \left(\frac{qe^{i\gamma}}{\sqrt{1-q^2}}\right)^{k_1} \left(\frac{qe^{-i\gamma}}{\sqrt{1-q^2}}\right)^{k_2} q^{2B}.
\end{equation}
Now consider the identical chord diagram, except for the fact that we have $k$ additional chords stretching between the two reservoirs. By \eqref{Hermite chord identity}, the contribution to $I_{n}(q,\gamma,n_F=4)$ is then:
\begin{equation}
        \frac{1}{[k]_{q^2}!} \left(\frac{q e^{i\gamma} qe^{-i\gamma}}{1-q^2}\right)^k \left(\frac{qe^{i\gamma}}{\sqrt{1-q^2}}\right)^{k_1} \left(\frac{qe^{-i\gamma}}{\sqrt{1-q^2}}\right)^{k_2}  q^{2B}.
\end{equation}
The contribution without reservoir to reservoir chords is thus modified by $\tfrac{\left(t_1 t_2\right)^k}{(1-q^2)^k[k]_{q^2}!}$
(we replaced $q e^{i\gamma} q e^{-i\gamma}$ with $t_1 t_2$ since the calculation holds for general values of the latter). 
This factor is independent of the specific chord diagram in question. The chords stretching from reservoir to reservoir factorize and contribute only a factor of: 
\begin{equation}
    \braket{\Psi_{\gamma^*}| \Psi_{\gamma}} = \sum_{k=0}^{\infty} \frac{1}{[k]_{q^2}!} \left(\frac{t_1 t_2}{1-q^2}\right)^k = \frac{1}{(t_1t_2;q^2)_\infty}.
\end{equation}
This agrees with $I_0(q,\gamma,n_F=4)$, which can be found (for example) in \cite{8cc0700ea80a492fa61a34ea71833640}.

\subsection{The Askey-Wilson picture}
\label{subsec:askey picture nF=4}

\begin{figure}[h]
    \centering
    \includegraphics[width=\linewidth]{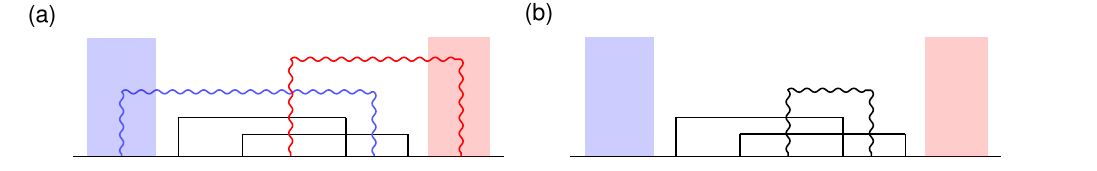}
    \caption{(a) A simple diagram with intersections between reservoir chords. (b) Replacing a pair of two reservoir chords with a special dynamical chord.}
    \label{fig:t2t1 diagram}
\end{figure}

The density of the $n_F=4$ Schur index \eqref{nF=4 Schur index} is the orthogonal measure for Al-Salam-Chihara polynomials $\varphi_n(\cos \theta;t_1,t_2|q^2)$ (see Appendix \ref{app: al salam}). We can define a continuum of states $\ket{\theta}_{n_F=4}$, satisfying  $\braket{n|\theta}_{n_F=4}=\varphi_n(\cos \theta;t_1,t_2|q^2)$ that are both orthonormal with respect to summation over $n$ and integration over $\theta$. The vector $\ket{\theta}_{n_F=4}$ then diagonalizes the transfer matrix:
\begin{equation}\label{nF=4 transfer matrix}
    T_{n_F=4}\ket{n} = \sqrt{[n+1]_{q^2}}\sqrt{1-t_1 t_2 q^{2n}} \ket{n+1} + \sqrt{[n]_{q^2}} \sqrt{1-t_1 t_2 q^{2n-2}} \ket{n-1} + \frac{t_1 + t_2}{\sqrt{1-q^2}} q^{2n} \ket{n}
\end{equation}
with eigenvalue $\frac{2\cos \theta}{\sqrt{1-q^2}}$. We can compute the vacuum to vacuum transfer matrix process:
\begin{equation}
    \begin{split}
        \bra{0} T_{n_F=4}^n \ket{0} & = \int_0^{\pi} d\theta \bra{0}T_{n_F=4}^n \ket{\theta}\braket{\theta|0} = \int_0^{\pi} d\theta |\varphi_0(\theta)|^2 \left(\frac{2\cos \theta}{\sqrt{1-q^2}}\right)^n \\[5pt]
        & = \frac{(t_1 t_2;q^2)_{\infty}}{2\pi} \int_0^{\pi} d\theta \frac{(q^2,e^{2\pm i \theta},q^2)_{\infty}}{(t_{1} e^{\pm i \theta},t_{2} e^{\pm i \theta};q^2)_{\infty}} \left( \frac{2\cos \theta}{\sqrt{1-q^2}} \right)^n \\[5pt] &  = (t_1 t_2;q^2)_{\infty} I_{n}(q,\gamma,n_F=4).
    \end{split}
\end{equation}
As for $n_F=2$, we would like to understand this transfer matrix in terms of the segment description. For that, it is useful to switch to a different basis $\{\ket{\bar{n}}\}$, such that the transfer matrix can be represented as:
\begin{equation}
    T_{n_F=4} \ket{\overline{n}} = \bigg(\mathfrak{a}^{\dagger}(1-t_1 t_2 q^{2\hat{n}}) + \mathfrak{a} + \frac{t_1 + t_2}{\sqrt{1-q^2}} q^{2\hat{n}} \bigg) \ket{\overline{n}}, \quad \ket{\overline{n}}=\sqrt{\frac{[n]_{q^2} !}{(t_1 t_2;q^2)_n}} \ket{n}.
\end{equation}
By $\mathfrak{a}$ and $\mathfrak{a}^{\dagger}$ we denote the operation,
\begin{equation}
    \mathfrak{a}\ket{\overline{n}} = [n]_{q^2} \ket{\overline{n-1}}, \quad \mathfrak{a}^{\dagger} \ket{\overline{n}} = \ket{\overline{n+1}}.
\end{equation}
$T_{n_F=4}$ is Hermitian as long as $t_1,t_2$ are complex conjugates of each other (or real separately), whereas
$\mathfrak{a}$ and $\mathfrak{a}^{\dagger}$ in this form are no complex conjugates of each other. 
To show how this transfer matrix arises from the segment description, we cut open the chord diagrams such that there is one reservoir to the left and one reservoir to the right, giving diagrams like in Figure \ref{fig:nF=4 transfer matrix}. The first reservoir is associated to a factor with $t_1$ and the second one with a factor of $t_2$. We call them $t_1$-reservoir and $t_2$-reservoir, respectively. As before, we distinguish between dynamical chords and $t_1$-, $t_2$-reservoir chords. The Hilbert space is spanned by dynamical chords only. At each step of the transfer matrix process, one of four things can happen:
\begin{enumerate}
    \item A dynamical chord is created.
    \item A dynamical chord is annihilated.
    \item A $t_1$-reservoir chord is annihilated.
    \item A $t_2$-reservoir chord is created. 
\end{enumerate}
For simplicity, we will for the moment rescale $t_{1,2} \rightarrow \sqrt{1-q^2} t_{1,2}$, such that each reservoir simply gives us a factor of $t_1$ or $t_2$. 
The crucial difference to the $n_F=2$ case is that the two different types of reservoir chords can now intersect and give rise to $q^2$ factors. Were this not the case, the transfer matrix:
\begin{equation} \label{almost correct transfer matrix}
    \tilde{T}_{n_F=4} = \mathfrak{a} + \mathfrak{a}^{\dagger} + (t_1 + t_2) q^{2n}
\end{equation}
would give the correct answer. Now we correct the remaining $q^2$ factors. Let us draw the chord diagrams in such a way that all the $t_2$-chords are above all the $t_1$-chords (see Figure \ref{fig:nF=4 transfer matrix}). Then a nontrivial crossing of reservoir chords happens every time a $t_1$-chord is annihilated after a $t_2$-chord is created. We would like to find the appropriate modification of the transfer matrix \eqref{almost correct transfer matrix}, that takes this into account. Looking at Figure \hyperref[fig:t2t1 diagram]{5}, we see that we can imitate the crossings of a pair of $t_1$-, $t_2$-chords by replacing it with a special dynamical chord that connects the $t_2$- with the $t_1$-nodes, and goes above all dynamical chords. A reasonable attempt would be to add to \eqref{almost correct transfer matrix} the following operator: 
\begin{equation}\label{TM modification operator}
    -t_1 t_2 (1-q^2) \mathfrak{a}^{\dagger} q^{2n}.
\end{equation}
This operator creates a special dynamical chord, which crosses all the other dynamical chords at creation due to the factor $q^{2n}$, and correctly accounts for the powers of $t_1$, $t_2$. We now show that \eqref{TM modification operator} is the correct deformation in two steps: in step 1, we prove that it gives the right result in the scenario where we only have reservoir chords. In step 2, we then show that diagrams with dynamical chords are also correctly taken into account, by proving that the special dynamical chord \eqref{TM modification operator} leads to the same crossing factors as the reservoir chords. 

\begin{figure}
    \centering
    \includegraphics[width=0.7\linewidth]{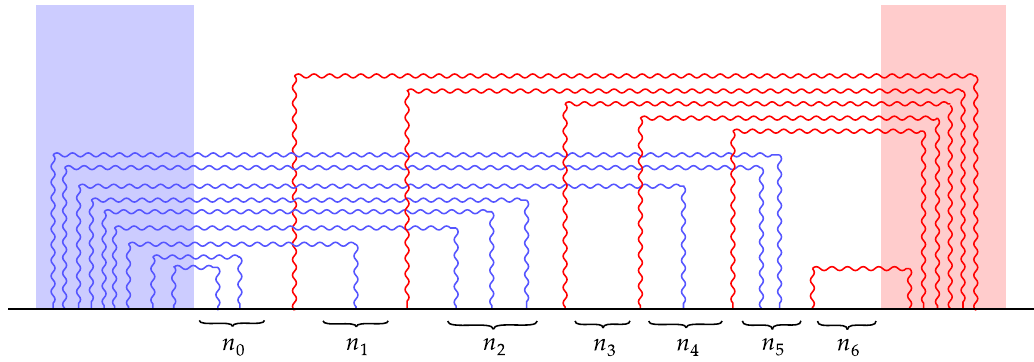}
    \caption{Example for a diagram with only reservoir chords. We think of the $t_2$-reservoir chords as partitioning the $t_1$-reservoir chords.}
    \label{fig:reservoir only}
\end{figure}

\paragraph{Step 1}
Consider a chord diagram built of only reservoir chords (see Figure \ref{fig:reservoir only}). We can think of the $t_2$-chords as partitioning the set of $t_1$-chords, with $n_i$ $t_1$-chords between the $i$-th and $i+1$-th $t_2$-chord. The total $q$-factor of such a diagram is then:
\begin{equation}\label{formula for crossings only reservoir}
    q^{2\sum_j j n_j}.
\end{equation}
Now assume that all the intersections have been taken into account correctly, except for the first $t_2$-chord. In such a case, we would get $q^{2\sum_j n_j (j-1)}$. At the first $t_2$-node, we now replace the $t_2$-chord with a dynamical chord by inserting \eqref{TM modification operator}. It can end on any of the $t_1$-nodes, except the $n_0$ $t_1$-nodes to the left. The first possibility is that it ends somewhere between the first and the second $t_2$-node. From counting crossings, we get:
\begin{equation}
    -(1-q^2)(1+q^2+...+q^{2(n_1-1)}) q^{2\sum_j n_j (j-1)} = -(1-q^{2n_1}) q^{2\sum_j n_j (j-1)}.
\end{equation}
The second possibility is that the dynamical chord ends somewhere in the region between the second and the third $t_2$-node. In this case, it picks up a factor:
\begin{equation}
    -(1-q^2) q^{2n_1+2} (1+q^2+...+q^{2n_2}) q^{2\sum_j n_j (j-1) - 2} = q^{2n_1} (1-q^{2n_2}) q^{2\sum_j n_j (j-1)}.
\end{equation}
The factor of $q^{2n_1+2}$ arises since the dynamical chord crosses all the $t_1$-chords between the first and the second $t_2$-node, and it also crosses the second $t_2$-node itself. The $-2$ in the last exponent is because we have one $t_1$-chord less that intersects the first and second $t_2$-chord. Similarly, for the dynamical chord ending between the $i+1$-th and $i+2$-th $t_2$-node, we have the contribution:
\begin{equation}
    -(1-q^2) q^{2\sum_{j=1}^i n_j + 2i} (1+q^2+...+ q^{2n_{i+1}} ) q^{2\sum_j n_j(j-1) - 2i} = -q^{2\sum_{j=1}^i n_j} (1-q^{2n_{i+1}}). 
\end{equation}
Summing over all these possibilities leads to a telescopic sum that collapses to:
\begin{equation}
    q^{2\sum_j n_j} q^{2\sum_j n_j (j-1)} = q^{2\sum_j j n_j },
\end{equation}
which proves step 1.

\paragraph{Step 2}
To see the general argument that the dynamical chords are correctly taken into account by the corrected transfer matrix \eqref{almost correct transfer matrix} $+$ \eqref{TM modification operator}, it is sufficient to consider a diagram where a single $t_1$-chord is annihilated to the right of where a single $t_2$-chord is created, with arbitrary dynamical chords. We will refer to such a diagram as $t_2 t_1$ diagram (see Figure \hyperref[fig:t2t1 diagram]{5(a)}).
We can replace the $t_1, t_2$ pair of reservoir chords with a special dynamical chord (squiggly black line in Figure \hyperref[fig:t2t1 diagram]{5(b)}) without changing the intersection factor, except for the crossing of the $t_1$- and $t_2$-chords themselves. Normally we have the prescription that when we create a dynamical chord it is below all the other chords. However, the special dynamical chord will be created on top of all the dynamical chords, as it originally replaces the pair of $t_1$-, $t_2$-chords. Therefore, the special dynamical chord is created by applying $t_1 t_2 a^{\dagger} q^{2n}$ at the $t_2$-node. After its creation, it is treated the same as all the other dynamical chords; hence the $t_1$-node can then simply replaced by an ordinary annihilation operator\footnote{Another way to see this is that on the one hand, the $t_1,t_2$ dependence is already exactly fixed by the deformation of the creation operator, and on the other hand the annihilation piece needs to be simply $a$ as $t_1,t_2\rightarrow 0$.}. In that way, the dynamical chords are agnostic as to whether they cross a $t_2 t_1$ configuration or a special dynamical chord. The same kind of argument works for arbitrary $t_1$- and $t_2$-chords. Replacing any $t_2$-node with the operator \eqref{TM modification operator} and $t_1$-nodes with annihilation operators only influences the crossings of reservoir chords.

Combining steps 1 and 2 and rescaling $t_{1,2}$ to their original value $t_{1,2}\to t_{1,2}/\sqrt{1-q^2}$, we have thus shown that the transfer matrix:
\begin{equation}
    T_{n_F=4} = \mathfrak{a} + \mathfrak{a}^{\dagger} (1-t_1 t_2 q^{2n}) + \frac{t_1 + t_2}{\sqrt{1-q^2}} q^{2n} 
\end{equation}
describes the Hilbert space of dynamical chords and agrees with the transfer matrix we get from the Askey-Wilson picture.

\subsection{Particle on the Quantum Disk}
\label{subsec:quantum disk}

The partition function of ordinary DSSYK is related to a boundary particle traveling on a noncommutative version of the hyperbolic disk \cite{Berkooz:2022mfk}. The particle starts off at some point on the boundary (corresponding to $\ket{0})$, travels through the noncommutative bulk and then goes back to the same point on the boundary. 

For the $n_F=4$ case, we instead compute $\bra{\Psi_{\gamma^*}} e^{-\beta T} \ket{\Psi_{\gamma}}$. In the noncommutative disk interpretation, as $\ket{\Psi_{\gamma}}$ is a coherent state with occupation number $\propto 1/\lambda$, the particle starts and ends in the bulk, at an order one distance away from the boundary. After an $SU_q(1,1)$ gauge transformation, the $n_F=4$ half-index should then correspond to the trajectories of a particle that starts and ends at the center of a noncommutative disk. There is a classical analogue of that computation: Kitaev and Suh \cite{Kitaev:2018wpr} computed the partition function of a particle on the ordinary hyperbolic disk with imaginary spin $-i\hat{\gamma}$. The density of states is different from a Schwarzian density (the latter is the $q\rightarrow 1$, low temperature limit of the $n_F=0$ index), but it is recovered in the limit of large spin. 

In Section \ref{subsubsec: particle on hpyerbolic disk}, we review the calculation of \cite{Kitaev:2018wpr} on the classical hyperbolic disk. In Section \ref{subsubsec: quantum disk}, we then recall some basics of quantum groups and the quantum disk.
In Section \ref{subsubsec: calculus on holomorphic line bundle}, we show that we can indeed compute the partition function of a particle on a specific noncommutative model of the hyperbolic disk (the quantum disk of \cite{Almheiri:2024ayc,vaksman2010quantumboundedsymmetricdomains}) to get the $n_F=4$ index. The low temperature limit where $\beta \gg 1-q^2$, $q \rightarrow 1$ recovers the density of states of Kitaev and Suh \cite{Kitaev:2018wpr}. The (imaginary) spin corresponds to the chemical potential in the $n_F=4$ index.

\subsubsection{The particle on the hyperbolic disk}
\label{subsubsec: particle on hpyerbolic disk}

The Schwarzian theory can be understood as a theory over cutouts of $AdS_2$ hyperbolic space specified by a curve $X$, with the following action\footnote{This action arises from the fact that the Schwarzian is roughly $K-1$, where $K$ is the boundary curvature, and the Gauss-Bonnet theorem \cite{Kitaev:2018wpr}.} \cite{Kitaev:2018wpr}:
\begin{equation}
    I_g[X]= - \hat{\gamma} (\text{area}[X]-L+2\pi), \quad \text{Length}[X]=L,
\end{equation}
where $\hat{\gamma}$ is a constant. The saddle point is given by the maximum of the area at given length, i.e. a circle. At large $\hat{\gamma}$ and $L$, this theory is quantum-mechanically equivalent to the Schwarzian theory\footnote{This explains a posteriori why we recover the Schwarzian density only in the large $\hat{\gamma}$ limit.} \cite{Kitaev:2018wpr}. Enforcing the length constraint with a Lagrange multiplier gives the action of a particle with periodicity or ``spin'' $\nu = -i\hat{\gamma}$. In order to find the Schwarzian partition function, we can then compute the partition function of a boundary particle on the disk, in the limit of large spin. The finite temperature partition function is found by summing over all closed paths of proper length $\beta$:
\begin{equation}
    Z(\beta) = \frac{1}{\text{Vol}(PSL(2,\mathbb R))} \int d^2 x \sqrt{g(x)} G(x,x;\beta) = \frac{1}{2\pi} G(0,0;\beta) = \frac{1}{2\pi} \int_0^{\infty} dE e^{-\beta E} |\psi_{E,\hat{\gamma}}(0)|^2.
\end{equation}
In the last step we have inserted an identity. The density of states is thus determined by the value of the wavefunctions at the origin. It is enough to consider $s$-wave functions, as others drop out of the calculation. The wave functions are found by solving the eigenvalue equation of the Casimir on the holomorphic line bundle, which encodes the spin of the particle. Practically, this amounts to writing down the Laplacian with spinor covariant derivatives. Going through the calculation, Kitaev and Suh \cite{Kitaev:2018wpr} observe the density of states:
\begin{equation}\label{Kitaev Suh density}
    \frac{1}{2\pi} |\psi_{E,\gamma}(0)| ^2=\rho(E) \propto \frac{\sinh{\sqrt{E}}}{\cosh{\sqrt{E}}+\cosh (2\pi \hat{\gamma})},
\end{equation}
which has a low-energy Schwarzian limit at large $\hat{\gamma}$.\footnote{Alternatively, we can think of the setup as a a particle in a large imaginary magnetic field \cite{Iliesiu:2019xuh}.} 

We formulate a q-analogue of that calculation, on a noncommutative version of the hyperbolic disk. In the following subsection, we set up the relevant notation and necessary tools to perform the q-analogue.

\subsubsection{The quantum disk}
\label{subsubsec: quantum disk}

We now give a very brief review of quantum group actions and the quantum disk as a representation of a quantum group. For details, we refer to Appendix \ref{sec: Quantum group basics}, and also \cite{Almheiri:2024ayc,vaksman2010quantumboundedsymmetricdomains}, whose notations we follow.

Consider first the ordinary hyperbolic disk, given by the metric:
\begin{equation}
    ds^2 = 4\frac{dz dz^*}{(1-zz^*)^2}, \quad |z z^*| \leq 1.
\end{equation}
It has an action of $\mathfrak{su}_{1,1}$ generators acting on it, given as:
\begin{equation}
    H = 2 z \partial_{z}-2z^* \partial_{z^*},\; E = -z^2 \partial_z + \partial_{z^*}, \; F = -(z^*)^2 \partial_{z^*} + \partial_z. 
\end{equation}
The generators fulfill a reality condition $E^*=-F$, $H^*=H$. The Lie algebra consisting of $E,F,H$ gives rise to a universal enveloping algebra $\mathcal{U}(\mathfrak{su}_{1,1})$. We want to define a noncommutative version of the disk and its algebra action. 
To make the disk noncommutative, a convenient choice turns out to be:
\begin{equation}\label{eqn:zzbar}
    z^* z = q^2 z z^* + 1-q^2,
\end{equation}
where $q$ is a number between zero and one. We want to find a corresponding deformation of the universal enveloping algebra  $\mathcal{U}_q(\mathfrak{su}_{1,1})$. It turns out that to do so, one needs to drop one of the Lie-algebra like elements and replace it with a group-like element. We have the generators $E,F$ (which are analogous to the Lie algebra elements), and the group-like invertible element $K$, which one can think of loosely as an exponentiated version of $H$. It is (an analytic continuation of) a finite rotation on the disk. The action on $z,z^*$ is:
\begin{equation}\label{quantum disk representation}
    \begin{split}
        & K(z) = q^2 z, \quad E(z) = -q^{\frac{5}{2}} z^2, \quad F(z) = q^{-\frac{3}{2}} \\
        & K (z^*) = q^{-2} z^*, \quad E(z^*) = q^{\frac{1}{2}}, \quad F(z^*) = -q^{\frac{1}{2}} (z^*)^2. 
    \end{split}
\end{equation}
Consistent with this action are the relations:
\begin{equation}
    KE = q^2 EK, \quad KF = q^{-2} FK, \quad EF - FE = \frac{K-K^{-1}}{q-q^{-1}}, \quad K K^{-1}=K^{-1}K=1.
\end{equation}
We still have to specify how these generators act on tensor products. This is done by the co-product:
\begin{equation}\label{coproduct}
    \Delta (E)= E \otimes 1 + K \otimes E, \quad \Delta(F)=F \otimes K^{-1} + 1 \otimes F, \quad \Delta(K)= K \otimes K. 
\end{equation}
One should read this equation as deformation of the usual Leibniz rule for vector fields. The algebra of $E,F,K,K^{-1}$ with these relations forms a $q$-deformed Hopf algebra, or a quantum group $\mathcal{U}_q(\mathfrak{su}_{1,1})$. It has a well-defined representation on the quantum disk given by \eqref{quantum disk representation}. For more details on the exact Hopf algebra structure, see Appendix \ref{sec: Quantum group basics}. The $q$-deformed algebra has a Casimir:
\begin{equation}
    C_q = EF + \frac{q^{-1}(K-1)+ q(K^{-1}-1)}{(q^{-1}-q)^2}, 
\end{equation}
which serves to construct a q-analogue of the Laplacian on the quantum disk. Finally, we consider differentiation and integration on the quantum disk. It is useful to introduce: 
\begin{equation}
    y = 1- zz^*,
\end{equation}
which commutes with $K$ and is thus a sensible radial coordinate. $y$ and $z,z^*$ do not commute, so the ordering is important in the following expressions. One can then put $z,z^*$ and $y$ as operators on a Hilbert space such that $y$ has discrete eigenvalues $q^{2n}$ \cite{Almheiri:2024ayc}. Then on functions $f(y)$, we can use the coproduct to show that $E,F$ have the following action \cite{vaksman2010quantumboundedsymmetricdomains}:
\begin{equation}\label{quantum group action on s-wave}
    E (f(y)) = -q^{\frac{1}{2}} zy \frac{f(y)-f(q^2y)}{y-q^2y}, \quad F (f(y)) = - q^{\frac{5}{2}}y \frac{f(y)-f(q^2y)}{y-q^2y} z^*.
\end{equation}
The fractions have the form of a finite difference operation. One can then define an invariant integral (see Appendix \ref{sec: Quantum group basics} for more details). For a function:
\begin{equation}
    f(z,z^*) = \sum_{n=0}^{\infty} z^n f_n(y) + f_0(y) + \sum_{n=0}^{\infty} (z^*)^n f_{-n}(y),
\end{equation}
we define the integration:
\begin{equation}\label{integration measure}
    \int_{\mathbb{D}_q} d\nu f(z,z^*) = \pi (1-q^2) \sum_{n=0}^{\infty} f_0(q^{2n}) q^{-2n}.
\end{equation}
This integration reduces to the ordinary hyperbolic integration in the case of $q\rightarrow 1$.
We can compute the Casimir operator on functions $f_0(y)$. It is \cite{vaksman2010quantumboundedsymmetricdomains}:
\begin{equation}
    C_q (f_0(y)) = - q \Box_q f_0(y),
\end{equation}
with $\Box_q$ being a $q$-deformation of the Laplacian on the hyperbolic disk:
\begin{equation}
    \Box_q f_0(y) = \mathcal{D}_q (x(1-q^{-1}x) \mathcal{D}_q)f_0(y),
\end{equation}
where $x=y^{-1}$ and: 
\begin{equation}
    \mathcal{D}_q f_0(x) = \frac{f_0(q^{-1}x)-f_0(qx)}{q^{-1}x-qx}.
\end{equation}
Finally, we discuss how to implement particles with ``spin'' on the quantum disk. For that, one must generalize the notion of a holomorphic line bundle to the quantum disk.

\subsubsection{Calculus on the holomorphic line bundle}
\label{subsubsec: calculus on holomorphic line bundle}

The calculus of holomorphic line bundles on the quantum disk has been explored in mathematical literature before \cite{vaksman2010quantumboundedsymmetricdomains} (see also \cite{5137fd7910a3479986566af4d366c58c}). The basic idea is to introduce the analogue of an Einbein that transforms correctly under $\mathcal{U}_{q}(\mathfrak{su}_{1,1})$ transformations. We introduce an object $v_{\mu}$ which obeys the relations \cite{vaksman2010quantumboundedsymmetricdomains}:
\begin{equation}
    z v_{\mu} = q^{-\mu} v_{\mu}z, \quad z^* v_{\mu} = q^{\mu} v_{\mu} z^*.
\end{equation}
The index $\mu$ denotes the spin of the particle, i.e. the eigenvalue under $K$ which is the q-analog of the rotation generator, and should not be confused with a spacetime index. 
We also have the action:
\begin{equation}\label{quantum group action on einbein}
    K v_{\mu} = q^{\mu} v_{\mu}, \quad Ev_{\mu} = -q^{\frac{1}{2}}\frac{1-q^{2\mu}}{1-q^2} zv_{\mu}, \quad F v_{\mu}=0.
\end{equation}
We can then understand the $s$-wave sections of the holomorphic line bundle as: 
\begin{equation}
    f(y) v_{\mu},
\end{equation}
with the action of $E,F,K$ governed by the rules \eqref{quantum group action on s-wave} and \eqref{quantum group action on einbein} and the coproduct \eqref{coproduct}. We can compute the Casimir (i.e. the Laplacian) on that representation:
\begin{equation}
    \begin{split}
        C_q(f(y) v_{\mu}) & = C_q(f(y)) q^{-\mu} v_{\mu} + f(y) \frac{1}{(q^{-1}-q)^2} (q^{\mu-1 }+ q^{1-\mu} - q - q^{-1}) v_{\mu} \\ &  + q \frac{f(y)-f(q^2 y)}{(1-q^2)^2}(1-q^2y) q^{-\mu} v_{\mu} (1-q^{2\mu}).
    \end{split}
\end{equation}
The spectrum of a spinor on the quantum disk can now be obtained by solving the eigenvalue equation $C_q(f(y) v_{\mu})=\lambda_{\mu} ( f(y) v_{\mu} )$. As $y$ takes values in $\{q^{2n}\}_{n \in \mathbb N_0}$, we can write $\phi_{n}=f(q^{2n})$. The eigenvalue problem then becomes:
\begin{equation}\label{holomorphic bundle recursion}
    \begin{split}
         \lambda_{\mu} \phi_n v_{\mu} = &  
         \phi_{n+1} v_{\mu} \frac{q}{(1-q^2)^2} q^{\mu} (1-q^{2+2n}) + \phi_{n-1} v_{\mu} \frac{q}{(1-q^2)^2} q^{-\mu} (q^2 -q^{2+2n})\\&  
         + \phi_n v_{\mu} \frac{q}{(1-q^2)^2} (q^{2n+2}(q^{-\mu} + q^{\mu}) - 1 -q^2).
    \end{split}
\end{equation}
The diagonal piece on the right hand side contains a $n$-independent constant that can be absorbed into $\lambda_{\mu}$. Then \eqref{holomorphic bundle recursion}, 
can be diagonalized by (rescaled) Al-Salam Chihara poylnomials, with analytic continuation $q^{\pm \mu}= t_{1,2} q^{-1} $. Explicitly, we have
\begin{equation}
    \phi_{n} = \,_3 \phi_2\bigg({q^{-2n}, q^{1-\mu}e^{\pm i\theta}\atop q^2, 0}; q^2,q^2\bigg),
\end{equation}
which is related to Al-Salam Chihara polynomials via \eqref{eqn:ASC}.

\noindent The energy spectrum of the Laplacian is:
\begin{equation}
    E(\theta)=\frac{1+q^2}{(1-q^2)^2}-\frac{2q\cos \theta}{(1-q^2)^2},
\end{equation}
where the $\theta$-independent term comes from the aforementioned constant in the diagonal piece of the recursion relation. 
Finally, the appropriate measure on the quantum disk for sections of the holomorphic bundle is given by \cite{vaksman2010quantumboundedsymmetricdomains}:
\begin{equation}
    \int_{\mathbb{D}_q} d\nu y^{\mu},
\end{equation}
where $d\nu$ is the integration measure defined in \eqref{integration measure}.
Normalizing the wave functions with respect to this integration gives the partition function, in exact analogy to the calculation of \cite{Kitaev:2018wpr}:
\begin{equation}\label{q disk final result}
    \begin{split}
         Z(\beta)& = \frac{1}{2\pi} \int dE\; e^{-\beta E} |\phi_0|^2 = \int_0^{\pi} \frac{d\theta }{2\pi} \frac{(q^2,q^2,e^{\pm 2i \theta};q^2)_{\infty}}{(q e^{\pm i \theta} q^{-\mu},q e^{\pm i \theta}q^{\mu};q^2)_{\infty}} e^{-\beta E(\theta)},
    \end{split}
\end{equation}
where we have normalized the partition function to $Z(0)=1$. After analytically continuing $q^{\mu} \rightarrow e^{i\gamma}$ (corresponding to imaginary spin), the density is exactly the $n_F=4$ Schur half-index density, with the coefficient of $\beta^n$ corresponding to index with $n$ fundamental Wilson lines inserted (up to some overall rescaling of the temperature). A particle on the quantum disk thus has the same partition function as the index of the $n_F=4$ theory. This means that the ground state of a particle on the quantum disk is a solitonic state in DSSYK. Taking the low temperature, $q\rightarrow 1$ limit, one can recover \eqref{Kitaev Suh density} from \eqref{q disk final result}.

\subsection{DSSYK, the quantum disk, and the bulk to boundary map}
\label{subsec: nF=4 discussion}
The quantum disk interpretation is related to the chord pictures discussed in Section \ref{sec:nF=4}:
\begin{equation}\label{bulk to boundary map}
    \text{Tr}(e^{-\#\beta H_{\text{Quantum Disk}}}) \cong \bra{0} e^{-\beta T_{n_F=4}} \ket{0} = (t_1 t_2;q^2)_{\infty} \bra{\Psi_{\gamma^*}} e^{-\beta H} \ket{\Psi_{\gamma}}.
\end{equation}
In the JT limit, the left hand side is a particle on the hyperbolic disk with the Kitaev-Suh density of states \eqref{Kitaev Suh density} (which reduces to the Schwarzian at large spin), while the right hand side is computed as a bulk JT amplitude between two wormholes (which are the EOW branes of \cite{Okuyama:2023byh}). The relation \eqref{bulk to boundary map} is the finite $q$ version of that statement: the left hand side is computed in the one-sided boundary Hilbert space, given by quantum mechanics of a particle on the quantum disk, while the right hand side can be interpreted to compute an amplitude between wormholes of finite length in the bulk \cite{Lin:2022nss}. The middle expression of \eqref{bulk to boundary map} has a double meaning and connects the two descriptions:  on one hand, we can understand it as the transfer matrix of dynamical chords, according to \eqref{subsec:askey picture nF=4}. On the other hand, there is an interpretation as a random walk on concentric circles \cite{Almheiri:2024ayc}. Rescaling $z,z^*$ by $\sqrt{1-q^2}$, we can rewrite \eqref{eqn:zzbar} as:
\begin{equation}\label{eqn:qDisk-qOsc}
    z\sim\sqrt{1-q^2} a^{\dagger},\;\; z^*\sim\sqrt{1-q^2} a \implies a a^{\dagger} -q^2 a^{\dagger} a =1.
\end{equation}
The coordinates of the quantum disk thus form a $q$-harmonic oscillator algebra\footnote{Of course, \eqref{eqn:qDisk-qOsc} is subtle, especially as $q\to 1$. For $q\neq 1$ it is just the statement that the abstract algebra generated by $z, z*$ corresponds to a q-harmonic oscillator algebra.}, with ground state $y=1$, i.e. the center of the disk, and excited states that are concentric circles with radius $1-q^{2n}$.  $\bra{0} e^{-\beta T} \ket{0}$ is then a random walk on these concentric circles, with initial and final value in the center. The transfer matrix dictates the probabilities of the random walk: at each step, we can either jump down or jump up one concentric circle, or we stay on the same circle. 

Microscopically, $T_{n_F=4}$ describes the dynamical chords of a DSSYK partition function with two extra reservoirs. The random walk on the noncommutative quantum disk can then be computed by chord diagrams. The center of the disk corresponds to a scenario with only reservoir chords. As we approach the boundary, the number of dynamical chords becomes large, recovering the semiclassical limit.

\section{Other matter}
\label{sec: other matter}
We briefly comment on the other possible matter additions for an $SU(2)$ theory that are asymptotically free or conformal: $n_F=6,8$, and a single adjoint hypermultiplet, which corresponds to $\mathcal{N}=2^*$, and in the massless case to the $\mathcal{N}=4$ theory. 

\subsection{\texorpdfstring{$n_F=6$}{}}
\label{subsec: nF=6}
For $n_F=6$, we have three hypermultiplets. By the general discussion in Section \ref{subsec:schur indices} halving the index breaks the flavor symmetry to an $SO(3)\sim SU(2)$; we only turn on a chemical potential for the Cartan of that group, parametrized by a phase $\gamma$. The Schur index with $n$ fundamental Wilson line insertions is then:
\begin{flalign}
\begin{aligned}
    \label{nF=6 schur index}
     I_{n}(q,\gamma,n_F=6) = \int_0^{\pi} \frac{d\theta}{4\pi} \sin^2\theta\;  \frac{(q^2,q^2e^{\pm 2i\theta};q^2)_{\infty}}{(qe^{\pm i\theta\pm i\gamma};q^2)_{\infty}(qe^{\pm i\theta};q^2)_{\infty}} \left(\frac{2\cos \theta}{\sqrt{1-q^2}} \right) ^n.
\end{aligned}
\end{flalign}
Only turning on an $SU(2)$ chemical potential also guarantees that the integrand is positive and that we can get a Hilbert space structure from this measure. 

\begin{figure}[H]
    \centering
    \includegraphics[width=\linewidth]{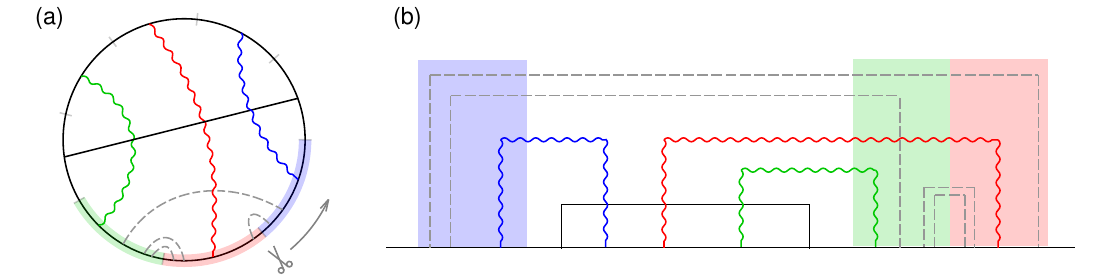}
    \caption{A chord diagram contributing to $\bra{0}T_{n_F=6}^5\ket{0}$. $t_1$-, $t_2$- and $t_3$-chords are drawn in blue, green and red, respectively. A new feature for $n_F\geq 6$ is that reservoir-to-reservoir chords (drawn in dashed gray lines) now intersect nontrivially with $t_i$-chords, hence they cannot simply be factored out.}
    \label{fig:nF=6 transfer matrix}
\end{figure}

\paragraph{The segment picture}
We sum over chord diagrams with three reservoirs, each of which is given by a coherent state with respect to the DSSYK q-annihilation operator. Dynamical complications arise in this case, because chords that go from reservoir to reservoir can interact nontrivially with dynamical chords (Figure \hyperref[fig:nF=6 transfer matrix]{6(a)}). In the $n_F=4$ case, this contribution factorizes, as chords going between reservoirs only change the overall normalization of $\ket{\Psi_{\gamma}}$. In this case, chords that go from reservoir to reservoir contribute more than just an overall factor. 

We can cut open the diagram such that one state segment sits at the beginning, and the other two segments sit at the end (Figure \hyperref[fig:nF=6 transfer matrix]{6(b)}). On the left, the in-state will simply be given by a coherent state. On the right, a chord going to each of the two reservoirs can be extracted by a factor of $\frac{q e^{i\pm \gamma}}{\sqrt{1-q^2}}$ as before, but now there will be extra $q$-factors from crossings between the chords. This gives rise to a $q^2$-binomial coefficient. The correct out-state is therefore a \textit{2-reservoir state}\footnote{In this formula, we have introduced a prefactor that will become clear from the segment description.}:
\begin{equation}\label{two reservoir state}
    \ket{\tilde{\Psi}_{\gamma_1,\gamma_2}} = \frac{1}{(qe^{i\gamma_1} qe^{i\gamma_2};q^2)_{\infty}} \sum_{n=0}^{\infty} \sum_{k=0}^n \binom{n}{k}_{q^2} \left( \frac{q e^{i\gamma_1}}{\sqrt{1-q^2}} \right)^k \left( \frac{q e^{i\gamma_2}}{\sqrt{1-q^2}} \right)^{n-k} \frac{1}{[n]_{q^2}!} \ket{\mathbf{n}}.
\end{equation}
In this case, the fugacities are $\gamma_1=\gamma$, $\gamma_2=-\gamma$. As before, the components are in terms of the chord basis $\ket{\mathbf{n}}$ \eqref{q Harmonic oscilator relations}.
To see how this state arises in the segment picture, we expand the matter contribution to the index:
\begin{equation}
    \frac{1}{(qe^{\pm i\theta};q^2)_{\infty} (qe^{\pm i\theta\pm i \gamma};q^2)_{\infty}},
\end{equation}
using \eqref{q Hermite generating functional}, as a sum over products of $q$-Hermite polynomials. Focusing just on the two ``right reservoirs'', we use the linearization formula\footnote{If one understands the $q$-Hermites as operators, then this formula is their fusion rule.} \cite{Ismail2005}:
\begin{equation}
    H_k(x|q^2) H_{k'}(x|q^2) = \sum_{m=0}^{min(k,k')} \frac{(q^2;q^2)_{k}(q^2;q^2)_{k'}}{(q^2;q^2)_{m}(q^2;q^2)_{k-m}(q^2;q^2)_{k'-m}} H_{k+k'-2m}(x|q^2).
\end{equation}
After some elementary manipulations, we find:
\begin{equation}
    \frac{1}{(qe^{\pm i \theta \pm i \gamma};q^2)_{\infty}} = \frac{1}{(qe^{i\gamma} qe^{-i\gamma};q^2)_{\infty}}\sum_{n=0}^{\infty} \sum_{k=0}^n \frac{\left(q e^{-i\gamma}\right)^k \left(q e^{i\gamma}\right)^{n-k}}{(q^2;q^2)_k (q^2;q^2)_{n-k}}  H_n(x|q^2).
\end{equation}
The formula holds identically for $\gamma \rightarrow \gamma_1$, $-\gamma \rightarrow \gamma_2$. 
From the $n_F=2$ discussion, we know that $H_n \leftrightarrow \sqrt{1-q^2}^n \ket{\mathbf{n}}$ and hence we recover \eqref{two reservoir state}. In summary, the state description for the $n_F=6$ transfer matrix is a propagation from a 2-reservoir to at 1-reservoir state \:
\begin{equation}
    I_{n}(q,\gamma,n_F=6) = \bra{\tilde{\Psi}_{\gamma,-\gamma}} H^n \ket{\Psi_0}. 
\end{equation}
Furthermore, any ordering of reservoirs will give the same answer, as the amplitude does not depend on how we cut open the chord diagram:
\begin{equation}
    I_{n}(q,\gamma,n_F=6) = \bra{\tilde{\Psi}_{\gamma,-\gamma}} H^n \ket{\Psi_0} = \bra{\tilde{\Psi}_{\gamma,0}} H^n \ket{\Psi_{-\gamma}} = \bra{\tilde{\Psi}_{-\gamma,0}} H^n \ket{\Psi_{\gamma}}.
\end{equation}
In a similar way, one should be able to write down an $n$-reservoir state, such that we could also express the half-index as an amplitude from a 3-reservoir to a 0-reservoir state, or more generally, from an $n_F/2-k$-reservoir state to a $k$-reservoir state. For $SU(2)$, having 2-reservoir states is sufficient, so we will not do so here. 

\paragraph{The Askey-Wilson picture}
Similar to the $n_F=2, 4$ cases, we can find a transfer matrix to describe the Schur half-index as a 1D process, now with three types of reservoir chords. As before, these matter contributions dress the creation and annihilation operators, and also contribute a diagonal term. The polynomials that are orthogonal with respect to the measure of \eqref{nF=6 schur index} are known as continuous dual $q$-Hahn polynomials (see Appendix \ref{app:qHahn}). We can write the eigenstates of $T_{n_F=6}$ in an orthonormal chord basis: 
\begin{flalign}
    \ket{\theta}_{n_F=6} = \sum_{n=0}^\infty \chi_{n}(\cos\theta;t_1,t_2,t_3|q^2)\ket{n},
\end{flalign}
where $\chi_{n}(\cos\theta;t_1,t_2,t_3|q^2)$ is given by \eqref{eqn:qHahn}. We reproduce \eqref{nF=6 schur index} by choosing $t_1=q, t_2=qe^{i\gamma}, t_3=qe^{-i\gamma}$, but we will keep these labels general, to better clarify the reservoir chord contributions. From the recursion relation \eqref{eqn:qHahn-recursion}, one then reads off the $T_{n_F=6}$ transfer matrix in the orthonormal basis:
\begin{equation}\begin{aligned}\label{eqn:nF=6-tmatrix}
    T_{n_F=6} =&
    \sqrt{\prod_{1\leq i<j\leq 3}(1-t_it_jq^{2n})} a^\dagger
    +\sqrt{\prod_{1\leq i<j\leq 3}(1-t_it_jq^{2n-2})} a\\&
    + \bigg(\frac{t_1+t_2+t_3}{\sqrt{1-q^2}}+\frac{t_1t_2t_3}{\sqrt{1-q^2}}(q^{2n-2}-q^{4n}-q^{4n-2})\bigg).
\end{aligned}\end{equation}
Where $a$ and $a^\dagger$ act as in \eqref{eqn:orthonormal basis}. By construction, $\ket{\theta}_{n_F=6}$ are eigenstates of this transfer matrix, with eigenvalues $\tfrac{2\cos\theta}{\sqrt{1-q^2}}$. Then:
\begin{equation}
    I_{k}(q,\gamma,n_F=6) = \frac{1}{(t_1 t_2;q^2)_{\infty}(t_1 t_3;q^2)_{\infty}(t_2 t_3;q^2)_{\infty}} \bra{0} T_{n_F=6}^k \ket{0},
\end{equation}
By setting one of the three $t_i\to 0$, we recover the $n_F=4$ transfer matrix \eqref{nF=4 transfer matrix}. Qualitatively, the $n_F=6$ transfer matrix \eqref{eqn:nF=6-tmatrix} shares the same features. The diagonal $t_1+t_2+t_3$ term creates reservoir chords, while factors of $(1-q^{2n}t_it_j)$ correct the creation/annihilation of dynamical chords. In fact, those contributions successfully describe all the chord diagrams with no mutually intersecting $t_1$-, $t_2$- and $t_3$-chords. 
The diagonal $t_1t_2t_3$ term corrects (in a subtle way) the crossing factors coming from chords that stretch from reservoir to reservoir\footnote{In the description as a state in the Hilbert space, these crossings are taken into account by the fact that we have a $q^2$-binomial in \eqref{two reservoir state}.}, as well as three mutually intersecting chords from different reservoirs. Presumably, one can write down a diagrammatic proof where one matches the transfer matrix exactly to the segment picture. We will not attempt to do so here. 

\subsection{\texorpdfstring{$n_F=8$}{}}
\label{subsec: nF=8}
For $n_F=8$, halving the index gives a $SO(4)\cong SU(2)\times SU(2)$ flavor symmetry. We can thus turn on fugacities parametrized by two phases $\gamma_1,\gamma_2$. The Schur index with $n$ fundamental Wilson line insertions is:
\begin{flalign}
    I_{n}(q,\gamma_1,\gamma_2,n_F=8) = \int_0^{\pi} \frac{d\theta}{4\pi} \sin^2 \theta \frac{(q^2,q^2e^{\pm 2i\theta};q^2)_{\infty}}{(qe^{\pm i\theta\pm i\gamma_1};q^2)_{\infty}(qe^{\pm i\theta\pm i \gamma_2};q^2)_{\infty}}  \left(\frac{2\cos \theta}{\sqrt{1-q^2}} \right) ^n.   
\label{nF=8 schur index}
\end{flalign}

\paragraph{The segment picture}
Similar to previous sections, we now sum over diagrams with four reservoirs, each of which represents as a coherent state of the q-annihilation operator. The picture is further complicated due to crossings between various reservoir-to-reservoir chords. 

For a state description, the essential conceptual jump was already performed for $n_F=6$. There, we wrote the Schur half-index as an amplitude between a 1-reservoir state and a 2-reservoir state. We can now simply replace the 1-reservoir state by another 2-reservoir state and to find: 
\begin{equation}
    I_{n}(q,\gamma_1,\gamma_2,n_F=8) = \bra{\tilde{\Psi}_{\gamma_1,-\gamma_1}} H^n \ket{\tilde{\Psi}_{\gamma_2,-\gamma_2}}.
\end{equation}
Notice that there is an exchange symmetry. We can fuse any of the two reservoirs into a 2-reservoir state.

\paragraph{The Askey-Wilson picture}
The polynomials that are orthogonal with respect to the measure of \eqref{nF=8 schur index} are Askey-Wilson polynomials \eqref{eqn:def-P}, with $t_1=t_2^*=qe^{i\gamma_1}$, $t_3=t_4^*=qe^{i\gamma_2}$. Each $t_i$ can be thought of as the coupling to a specific reservoir. We normalize energy eigenstates so that they are $\delta$-normalizable, with respect to the unit measure $d\theta$:
\begin{equation}
    \ket{\theta}_{n_F=8}=\sum_{n=0}^\infty \pi_n(\cos\theta;t_1,t_2,t_3,t_4|q^2)\ket{n}.
\end{equation}
We can use the recursion relation for Askey-Wilson polynomials \eqref{eqn:AW_tmatrix} to construct a chord transfer matrix $T_{n_F=8}$.
The Schur half-indices with $k$ fundamental Wilson lines can then be written as vacuum amplitudes in the chord Hilbert space:
\begin{equation}
    I_{k}(q,\gamma_1,\gamma_2,n_F=8) = \frac{(t_1t_2t_3t_4;q^2)_{\infty}}{(t_1t_2,t_1t_3,t_2t_3,t_1t_4,t_2t_4,t_3t_4;q^2)_{\infty}} \bra{0} T^k_{n_F=8} \ket{0}. 
\end{equation}
We have inserted the normalization factor according to the $k=0$ result \cite{8cc0700ea80a492fa61a34ea71833640}, but could have also computed it from the overlap of the 2-reservoir states.
The transfer matrix $T_{n_F=8}$ looks very tedious. By setting specific $t_i$'s to zero, one can check that it reduces to the $n_F<8$ cases, discussed previously. Each term in $T_{n_F=8}$ loosely plays a similar role (creating/annihilating dynamical and reservoir chords, correcting for intersections of reservoir chords), however, we leave a detailed understanding of this transfer matrix to future work.

\subsection{Adjoint Matter}\label{subsec: adjoint matter}
Finally, we consider $\mathcal{N}=2$ $SU(2)$ gauge theories with one adjoint hypermultiplet. For more than one the theory is infrared-free, hence we only study the case of a single adjoint hypermultiplet, which corresponds to $\mathcal{N}=2^*$ $SU(2)$ SYM. For an adjoint matter hypermultiplet, the single letter partition function is given by \eqref{hypermultiplet single letter - adj}:
\begin{equation}
    f^{H,adj}(q^2) = \frac{2q}{1-q^2}\chi_1(\theta), \quad \chi_1(\theta) = e^{2i\theta} + 1 + e^{-2i\theta}.
\end{equation}
The Schur half-index in the background of $n$ fundamental Wilson lines therefore reads:
\begin{equation}
    I_n^{adj}(q) = \int_0^{\pi} \frac{d\theta}{4\pi} \sin^2 \theta (q^2,q^2e^{\pm 2i\theta};q^2)_{\infty} \frac{1}{(qe^{\pm 2i\theta};q^2)_{\infty} (q;q^2)_{\infty}} \left(\frac{2\cos \theta}{\sqrt{1-q^2}}\right)^n.\label{eqn:adjoint1}
\end{equation}
We now present two ways of understanding this object.

\paragraph{The segment picture.}
The matter contribution to the index can be rewritten as a bilinear sum of q-Hermite polynomials, by using the identity \eqref{eqn:bilin-qH-gf}, for $t=q$:
\begin{equation}
    \frac{1}{(q;q^2)_{\infty} (qe^{\pm 2i\theta};q^2)_{\infty}} = \frac{(q;q^2)_{\infty}}{(q^2;q^2)_{\infty}} \sum_{p=0}^{\infty} H_p(x|q^2) H_p(x|q^2) \frac{q^p}{(q^2;q^2)_p}.\label{eqn:adjoint2}
\end{equation}
Hence it is natural to understand the half-index as a sum over chord diagrams with two reservoirs. Crucially, this sum does not factorize and the number of chords emanating from the two segments are constrained to be equal. The two reservoirs form an entangled state. 

\noindent By cutting open the chord diagrams, the resulting process resembles an evolution with respect to a density matrix  $\rho$. Expanding $\rho$ in the orthonormal basis, the $n$-th Hamiltonian moment in the mixed ensemble is given by:
\begin{equation}
    \text{Tr}(\rho\, H^n) = \sum_{k,l=0}^{\infty}\rho_{k,l} \bra{l} H^n \ket{k} = \sum_{k,l=0}^{\infty}\rho_{k,l}\int_0^{\pi} d\theta \braket{l|\theta} \braket{\theta|k} \left(\frac{2\cos \theta}{\sqrt{1-q^2}}\right)^n.
\end{equation}
Using the orthonormalized wavefunctions for $\braket{k|\theta}\equiv \psi_k(\cos\theta;q^2)$, defined in \eqref{eqn:ort-qH}, and the relation $(e^{\pm 2i\theta};q^2)_\infty = 4\sin^2\theta (q^2e^{\pm 2i\theta};q^2)_\infty$, we rewrite the trace as:
\begin{equation}
    \sum_{k,l=0}^{\infty}\rho_{k,l}\int_0^{\pi} d\theta\; 4\sin^2 \theta\; (q^2,q^2e^{\pm 2i\theta};q^2)_{\infty} \frac{H_k(\cos\theta|q^2) H_{l}(\cos\theta|q^2)}{2\pi\sqrt{(q^2;q^2)_k (q^2;q^2)_{l}}} \left(\frac{2\cos \theta}{\sqrt{1-q^2}}\right)^n.
\end{equation}
By matching to the half-index density \eqref{eqn:adjoint1}-\eqref{eqn:adjoint2}, we find a mixed (even diagonal) state in the orthonormal number basis:
\begin{equation}
    \rho = \sum_{k=0}^\infty \frac{1}{8} q^k \frac{(q;q^2)_{\infty}}{(q^2;q^2)_{\infty}} \ket{k}\bra{k}.
\end{equation}
This interpretation stands in contrast to cases with fundamental matter, where we interpret the Schur half-index as a transition amplitude between two (pure) $n$-reservoir states.

\paragraph{The Askey-Wilson picture.}
In analogy to the previous sections, the Schur half-index of $N=4$ $SU(2)$ SYM can also be rewritten as a vacuum-to-vacuum amplitude of a generalized, SYK-like model. In this case, the states in the chord basis are written in terms of continuous q-ultraspherical polynomials $C_n(\cos\theta;\beta|q^2)$, which also fall under the q-Askey scheme (see Appendix \ref{sec:qUltraspherical}).

Like in the $n_F=4$ case, discussed in Section \ref{sec:nF=4}, the segment picture here contains two reservoirs. However, as the number of chords from each reservoir constrained to be the same, the number of parameters effectively reduces to a single real variable $\beta$, which we can think of as the factor for creating reservoir chords. To reproduce the Schur half-index density of \eqref{eqn:adjoint1}, we need to take $\beta=\sqrt{q}$. We will still use $\beta$ for the remainder of this section, keeping the notation general.

For a 1D interpretation, it is convenient to rescale the energy eigenstates as $\ket{\theta}_{adj}=\sum_{n=0}^\infty \zeta_n(\cos\theta;\beta|q^2)\ket{n}$, so that they are $\delta$-normalizable, with unit measure (see \eqref{eqn:zeta_n}). Schur half-indices with $k$ fundamental Wilson lines can then be written as moments in the chord Hilbert space:
\begin{flalign}
    I_k^{adj}(q) = 
    \frac{(\beta^2q^2;q^2)_{\infty}}{(\beta,\pm \beta q,\pm \beta q,-\beta q^2;q^2)_\infty}
    \bra{0}T_{adj}^k\ket{0},
\end{flalign}
where the adjoint transfer matrix $T_{adj}$ can be read off from the recursion relation \eqref{eqn:zetaRR}. It does not contain any diagonal terms, unlike the cases with fundamental matter. We leave a more thorough analysis of $T_{adj}$ to future work.

\section{Outlook}
\label{outlook}

The SYK model is usually associated with the study of quantum chaos. As such it has little to do with the BPS sector of ${\cal N}=2$ theories, which is as ordered a problem as it gets. Understanding why the two are related is an interesting question. A recent idea in that general direction is the notion of BPS chaos \cite{Lin:2022rzw,Lin:2022zxd,Chen:2024oqv} when the coupling is taken to be finite. Some other issues worth exploring are the following: 
\begin{itemize}[topsep=1em,itemsep=0em]
    \item Another promising place to look for a connection to chaos is in the near-BPS sector. The same non-commutative algebra also appears in the discussion of stationary solutions of ASEPs which are related to classical probabilistic processes \cite{Derrida:1992fi, Sasamoto_2000, Golinelli_2006}. If we denote a space of configurations in an ASEP by $x$, and $P(x)$ is the probability measure on that space, then the ASEP defines the function $F$ in the equation ${\dot P}(x)=F(x)$. Stationarity then implies ${\dot P}(x)=F(x)=0$. If we think about $F$ as a Hamiltonian and $P(x)$ as a wave function, then the relation $F(x)=0$ is not that different from a BPS condition. To see whether there is such a relation one will need to go to the near BPS sector, which might be doable in this case.
    
    \item With the methods of \cite{Lewis:2025qjq}, one can attempt to write down a chord diagram description of the Schur half-index for $SU(N)$ gauge theories. For example, using the generating function for $q$-Whittaker polynomials, one can show that the $SU(N)$ $\mathcal{N}=4$ Schur half-index is a sum over chord diagrams, analogous to the $SU(2)$ case discussed in Section \ref{subsec: adjoint matter}. Similarly, one can speculate about more complicated representation theory structures. For example, from the Feynman rule description for $n$-point functions in DSSYK \cite{Berkooz:2018jqr}, one can read off that an $SU(2)$ bifundamental hypermultiplet corresponds to a matter chord of dimension $\Delta = \frac{1}{2}$. It would be interesting to see whether the Schur half-indices of all class $\mathcal{S}$ theories can be formulated as $n$-point functions of matter operators in some (generalized) DSSYK system, and what that might teach us about $\mathcal{N}=2$ gauge theories.

    \item For $n_F=4$, we have seen that we have to supplement our chord diagrams with two reservoir regions. In terms of the chord Hilbert space, we computed an amplitude between two coherent states whose occupation number scales as $n \sim \frac{1}{\lambda}$. This is exactly the limit in which the fake disk algebra of \cite{Lin:2023trc} arises. Insertions of matter operators (in the DSSYK sense) cannot happen in the reservoir region, and in that sense, they resemble insertions in the ``physical'' region in the fake disk picture. One can therefore speculate whether the two ``fake'' segments in the fake disk are the reservoir segments. In the semiclassical limit, one can compute the renormalized length of the reservoir segments and finds them to be zero, which is in agreement with the fake disk picture. On the other hand, since the $n_F=4$ case is related to the quantum disk, one can speculate about a connection between the quantum disk and the fake disk.
\end{itemize}

\acknowledgments
We thank Federico Ambrosino, Anirudh Deb, and especially Ohad Mamroud for discussions. JS would like to thank the University of Boulder Colorado and the Institute for Advanced Study for their hospitality while part of this work was carried out. The work of MB, TK and JS is supported in part by the Israel Science Foundation grant no. 2159/22, by the Minerva foundation, and by a German-Israeli Project Cooperation (DIP) grant "Holography and the Swampland". MB is an incumbent of the Charles and David Wolfson Professorial Chair of Theoretical Physics. The work of JS is supported by a research grant from the Chaim Mida Prize in Theoretical Physics at the Weizmann Institute of Science, and by a Dean Award of Excellence of the Weizmann School of Science.

\appendix

\section{Quantum group basics}
\label{sec: Quantum group basics}
Here, we briefly explain basics of quantum groups. We refer to the papers \cite{Almheiri:2024ayc,Berkooz:2022mfk} and the book\footnote{The PDF in the arXiv is in Russian. We thank Fedor Popov for providing us with a copy of a translation into English.} \cite{vaksman2010quantumboundedsymmetricdomains} for more thorough discussions.

Classical groups are a mathematically very rigid structure: they come in four discrete series, and a few exceptional ones. On the other hand, in a ``quantum'' deformation of a group, we would like to introduce a parameter $q = e^{-\# \hbar}$, and take the parameter $\hbar \rightarrow 0$ when we want to see the classical limit. We therefore need to have a structure with a continuous parameter $q^2$. The group structure directly is not deformable in such a way, but we can deform two related structures: the algebra of functions on the group, and the universal enveloping algebra of the Lie algebra. The deformations of both of these structures are dual to each other, as they are both Hopf algebras.

\paragraph{Algebra of functions on a group}

It is the viewpoint of algebraic geometry that one can study an object by studying properties of the algebra of functions on the object. In view of this idea, we can define a \textit{quantum group} by $q$-deforming the algebra of functions on the group, and working with that algebra instead of the (non-existing) underlying object. Ordinary functions on a group form a so-called Hopf algebra, as we will review now. 

\noindent Consider the example of $SU(1,1)\cong SL(2,\mathbb{R})$. We can represent it as $2\times 2$ matrices:
\begin{equation}
    SU(1,1)\cong SL(2,\mathbb R) = \left\{ \begin{pmatrix}
        a & b\\
        c & d
    \end{pmatrix}\Bigg| ad-bc=1 \right\}.
\end{equation}
$a,b,c,d$ are therefore like coordinates on the group manifold. Any (complex) function on a group can be approximated to arbitrary accuracy by a polynomial in the letters $a,b,c,d$, where we identify the polynomials which can be mapped to each other by commutation of the letters and the condition $ad-bc=1$. The algebra of functions on $SU(1,1)$ is given by:
\begin{equation}
    \mathcal{O}(SU(1,1))=\mathbb{C}[a,b,c,d]/\sim,
\end{equation}
where $\sim$ is the equivalence relation generated by the ideal $ab=ba, ..., ad=da, ad-bc=1$. $\mathcal{O}(SU(1,1))$ is clearly a commutative algebra as we can add and multiply functions together. There is also a neutral element: the function that is one everywhere. We can think of multiplication as a map:
\begin{equation}
    \begin{split}
        \cdot : & \mathcal{O}(SU(1,1))\otimes \mathcal{O}(SU(1,1)) \rightarrow \mathcal{O}(SU(1,1)) \\
        & (f_1 \cdot f_2)(g) = f_1(g) f_2(g).
    \end{split}
\end{equation}
Sometimes it is useful to denote the product as $m$ instead of the $\cdot$ (i.e. $m(f_1,f_2)=f_1 \cdot f_2$). 
We will often drop both the $\cdot$ and the $m$.
The group structure of $SU(1,1)$ allows us to establish a Hopf algebra structure. We define\footnote{Here we have assumed that $\mathcal{O}(SU(1,1))\otimes \mathcal{O}(SU(1,1)) \cong \mathcal{O}(SU(1,1)\times SU(1,1))$, which is true for compact groups, but subtle in the noncompact case.}:
\begin{equation}
    \begin{split}
        & \eta: \mathcal{O}(SU(1,1)) \rightarrow \mathbb C, \quad \eta(f)= f(e) \\
        & S: \mathcal{O}(SU(1,1)) \rightarrow \mathcal{ O}(SU(1,1)), \quad (Sf)(g)= f(g^{-1}) \\
        & \Delta: \mathcal{O}(SU(1,1)) \rightarrow \mathcal{O}(SU(1,1)) \otimes \mathcal{O}(SU(1,1)), \quad \\
        & (\Delta f)(g_1,g_2) = g(g_1 g_2). 
    \end{split}
\end{equation}
$\Delta$ is known as the co-product, $\eta$ as the co-unit (they together form a co-algebra structure), and $S$ is the antipode.
The group axioms then imply the identities:
\begin{equation}\label{Hopf algebra identities}
    \begin{split}
        & (\Delta \otimes id) \Delta = (id\otimes \Delta) \Delta \quad \text{(co-associativity)} \\
        & (id \otimes \eta)\Delta = (\eta \otimes id)\Delta = id \\
        & \Delta(f_1 f_2)=\Delta(f_1)\Delta(f_2), \, \eta(ab)=\eta(a)\eta(b) \quad \text{(bi-algebra)} \\
        & m(S\otimes id)\Delta = m(id\otimes S) \Delta = \eta \quad \text{(Hopf algebra)}.
    \end{split}
\end{equation}
We can work out the explicit form of these maps on the generators $a,b,c,d$:
\begin{equation}
    \begin{split}
        & \Delta(a) = a \otimes a + b \otimes c \\
        & \Delta(b) = a \otimes b + b \otimes d \\
        & \Delta(c)= c \otimes a + d \otimes c \\
        & \Delta(d) = d \otimes d + c \otimes b \\
        & \eta(a)=\eta(d)=1, \, \eta(b)=\eta(c)=0 \\
        & S(a)=d,\,S(b)=-c\,S(c)=-b, S(d)=a .
    \end{split}
\end{equation}

This is the structure of a Hopf algebra. For $\mathcal{O}_q(SU(1,1))$, the Hopf algebra is commutative. We thus have the rough correspondence:
\begin{equation}
    \text{classical group} \sim \text{commutative Hopf algebra of functions}.
\end{equation}
Quantum mechanics is essentially the study of noncommutative algebras. To get a quantum group, we deform the right hand side to a noncommutative algebra. For $SU(1,1)$, there is a unique $q$-deformation, which we call $\mathcal{O}_q(SU(1,1))$. It is defined as\footnote{Technically, this defines $\mathcal{O}_q(SL(2,\mathbb{C})).$ We still need to define a reality condition to get to $\mathcal{O}_q(SU(1,1))$, which we will handle in a moment.}:
\begin{equation}
    \begin{split}
        & \mathcal{O}_q(SU(1,1)) = \mathbb{C}[a,b,c,d]/I,\\
        I = (& ab-qba, ac-q ca, bd-q db, cd-q dc, \\ & bc-cb, ad-da-(q-q^{-1})bc, ad-qbc-1).
    \end{split}
\end{equation}
$I$ is the ideal which generates all the relations we quotient $\mathbb C[a,b,c,d]$ by. We then need to modify the Hopf algebra structure such that it is compatible with the ideal. We get that $\Delta, \eta$ act identical to before, while:
\begin{equation}
    S(a)= d,\, S(b)=-q^{-1} c,\, S(c)=-q b, \, S(d)=a.
\end{equation}
Finally, let us discuss the reality condition. Classically, we can get from $SL(2,\mathbb C)$ to $SU(1,1)$ by demanding $d^*=a, b^*=c$. For the quantum group, we have a $\ast$-involution generated by:
\begin{equation}
    a^*=d, \; b^* = q c.
\end{equation}
This is one view on quantum groups; there is a dual picture in terms of a $q$-deformation of the universal enveloping algebra, which we will discuss now.\\[5pt]

\paragraph{Universal enveloping algebra}
Instead of functions on the group itself, we can also consider a dual picture, given by some (linear) action on the algebra of functions. One such action is given by directional derivatives of vector fields acting on the functions. Since such an action should also encode the group structure in some way. As Lie algebra elements are in one to one correspondence with left-invariant vector fields, we can think of the \textit{universal enveloping algebra} as acting on the algebra of functions. We define\footnote{Once again up to a reality condition that we will define momentarily.}:
\begin{equation}
    \mathcal{U}(\mathfrak{su}_{1,1}) = \mathbb{C}[E,F,H]/\sim,
\end{equation}
where $E,F,H$ are the generators of the $\mathfrak{su}_{1,1}$ algebra and $\sim$ is the equivalence relation generated by the ideal $[E,F]=H, [H,E]=2E, [H,F]=-2F$. This set is clearly an algebra. It is actually also a Hopf algebra. We can define the mappings:
\begin{equation}
    \begin{split}
        & \Delta(H) = 1 \otimes H + H \otimes 1 \\
        & \Delta(E) = 1 \otimes E + E \otimes 1 \\
        & \Delta(F) = 1 \otimes F + F \otimes 1 \\
        & S(H)=-H,\;S(E)=-E,\; S(F)=-F \\
        & \eta(H)=\eta(E)=\eta(F) = 0.
    \end{split}
\end{equation}
Note that the co-product is just the ordinary Leibniz rule for taking  derivatives. One can check that the identities \eqref{Hopf algebra identities} are fulfilled. $U(\mathfrak{su}_{1,1})$ is thus also a Hopf algebra. It is very intuitive to think of this algebra as encoding (most of) the information about the group, as we can think of group elements as exponentials of the Lie algebra elements. 
For a generic element that is a polynomial in $E,F,H$, we can define the action on a function $f$ by consecutive application:
\begin{equation}
    E^n F^m H^k (f) = E(...(F(...(H(...Hf)...)...)...)...).
\end{equation}
We can evaluate the resulting function at the identity. This gives a bracket:
\begin{equation}
    \begin{split}
        & \langle \cdot, \cdot \rangle : \mathcal{U}(\mathfrak{su}_{1,1})\otimes \mathcal{O}(SU(1,1)) \rightarrow \mathbb{C}\\
        & \langle H, f\rangle \equiv (Hf)(e).
    \end{split}
\end{equation}
One can show \cite{Almheiri:2024ayc} that this bracket is compatible with both algebra structures, making the two algebras dual to each other. Therefore, loosely:
\begin{equation}
    \text{classical group} \sim \text{universal enveloping algebra}.
\end{equation}
We can $q$-deform the universal enveloping algebra. This can be done in such a way that there still exists a pairing, i.e. they continue to be dual to each other. 

The main modification is that instead of having only algebra-like elements in the generating set of the algebra, we have also group-like elements that are invertible. Concretely, we define the $q$-deformed universal enveloping algebra as:
\begin{equation}
    \mathcal{U}_q(\mathfrak{su}_{1,1}) \equiv \mathbb C[E,F,K,K^{-1}]/\sim,
\end{equation}
where $\sim$ is an equivalence relation generated by:
\begin{equation}
    EF-FE = \frac{K - K^{-1}}{q-q^{-1}},\; KE-q^2 EK =0,\; KF - q^{-2} FK=0.
\end{equation}
The way to think about these equivalence relations is to think about $K$ as ``$K = e^{2\log q H}$''. Correspondingly, the Hopf algebra structure gets modified. We have that:
\begin{equation}
    \begin{split}
        & \Delta (K) = K \otimes K \\
        & \Delta(E) = E \otimes 1 + K \otimes E \\
        & \Delta(F) = F \otimes K^{-1} + 1 \otimes F\\
        & \eta(K)=1,\; \eta(F)=\eta(E)=0 \\
        & S(K) = K^{-1},\;  S(K^{-1})= K, \; S(E)=-K^{-1} E, \; S(F)=-FK.
    \end{split}
\end{equation}
We can understand the co-product for $E,F$ as twisted Leibniz rule, which has also appeared in a different context in \cite{Lin:2023trc}.
Finally, the reality condition is the involution:
\begin{equation}
    (K^{\pm 1})^* = K^{\pm 1}, \quad E^*=-KF, \quad F^* = -E K^{-1}.
\end{equation}
One can find a Casimir operator that commutes with $E,F,K$, it being:
\begin{equation}
    C_q = EF + \frac{q^{-1}(K-1)+ q(K^{-1}-1)}{(q^{-1}-q)^2}.
\end{equation}

\section{Special functions and the q-Askey scheme}
\label{sec: q-Askey scheme}

In this Appendix, we list some basic notions of $q$-combinatorics. We define the q-Pochhammer symbol and the usual shorthand for a product thereof:
\begin{flalign}
    &(a;q^2)_n = \prod_{k=1}^n (1-a q^{2(k-1)}),\\&
    (a_1,a_2,..,a_r;q^2)_n = (a_1,q^2)_n ... (a_r,q^2)_n,\\[10pt]&
    (ae^{\pm i\theta};q^2)_n = (e^{i\theta};q^2)_n(e^{-i\theta};q^2)_n.
\end{flalign}
Throughout the text we often use q-numbers and q-factorials, defined as follows:
\begin{flalign}
    &[n]_{q^2} = \frac{1-q^{2n}}{1-q^2},\\&
    [n]_{q^2}! = [n]_{q^2}[n-1]_{q^2}...[1]_{q^2} = \frac{(q^2;q^2)_n}{(1-q^2)^n},
\end{flalign}
as well as the q-analog of the multinomial coefficient, and exponential function:
\begin{flalign}
    \binom{n}{a_1,a_2,...a_r}_{q^2} =& \frac{[n]_{q^2}!}{[a_1]_{q^2}!...[a_r]_{q^2}!}, \text{ where }
    \sum_{i=1}^r a_i = n. \label{eqn:qmulti}\\[4pt]
    \text{exp}_{q^2}(x) =& \sum_{k=0}^\infty \frac{x^k}{[k]_{q^2}!}\label{eqn:qexp}
\end{flalign}
\noindent When computing supersymmetric indices, one typically starts with a  single-letter particle partition function $f(q,z)$, we obtain the multi-particle contribution to the index by taking the plethystic exponential:
\begin{flalign}\label{eqn:plethystic}
    PE[f(q,z)]\equiv \text{exp}\bigg[\sum_{n=1}^\infty \frac{1}{n}f(q^n,z^n)\bigg].
\end{flalign}
Often, the plethystic exponential is a rational function, and working it out gives a product over $q$-Pochhammers. \\

In the main text we make use of various basic hypergeometric polynomials, which organize themselves into a hierarchy, described by the q-Askey scheme \cite{Koekoek2010,Ismail2005,Gasper2004}. Such polynomials commonly appear in literature on exactly solvable statistical mechanics models, notably asymmetric exclusion processes (ASEP) \cite{Uchiyama_2004, Sasamoto_2000, Watanabe:2024vad, Wang_2024}. They also frequently arise in the context of DSSYK and representation theory of $\mathcal{U}_q(\mathfrak{su}_{1,1})$ \cite{Berkooz:2022mfk,Mertens:2017mtv,Gabai:2024puk,Gabai:2024qum}.

The scheme can essentially be understood as a bounded four-dimensional space described by coordinates $(t_1,t_2,t_3,t_4)$ \cite{Koekoek2010}. We will only be interested in the case where $|t_i|\leq 1$. At each point in this space sits an infinite family of polynomials\footnote{For our purposes, we will replace the q-deformation parameter in $p_n(x;t_1,t_2,t_3,t_4|q)$ by $q^2$ throughout.} $p_{\{n\}}(x;t_1,t_2,t_3,t_4|q^2)$ indexed by their degree $n$. The most general polynomials in the q-Askey scheme are called Askey-Wilson polynomials, and are characterized by all $t_i\neq 0$.

Our aim is to show that polynomials of the q-Askey scheme diagonalize the various transfer matrices encountered in the main text. We do so by interpreting the action of each transfer matrix as the recurrence relation of the corresponding family of polynomials. In the process, we normalize the chord number states so that they form an orthonormal basis. For each family of polynomials, we recount the three relevant properties, which are orthogonality in $\theta$ and $n$, and the recursion relations\footnote{Although the simpler families of polynomials may be viewed as a special cases of the most general Askey-Wilson polynomials $p_{\{n\}}(x;t_1,t_2,t_3,t_4|q^2)$, we derive their orthogonality relations separately. This is because the derivation requires taking limits which may generically change the support or introduce singularities in the orthogonality measure.}. \\

\noindent All polynomials of the q-Askey scheme can be represented in terms of the basic hypergeometric series \cite{Koekoek2010}:
\begin{flalign}\label{eqn:hypergeometric}
    _r\phi_s\bigg({a_1,...,a_r\atop b_1,...,b_s}; q,z\bigg) = 
    \sum_{k=0}^\infty \frac{(a_1,...a_r;q)_k}{(b_1,...b_s;q)_k}(-1)^{(1+s-r)k} q^{(1+s-r)\binom{k}{2}} 
    \frac{z^k}{(q;q)_k}.
\end{flalign}

\noindent We start from the simplest case, corresponding to pure $SU(2)$ gauge theory and go over to cases involving fundamental hypermultiplets. Lastly, we discuss the case of a single adjoint hypermultiplet, i.e.  $\mathcal{N}=4$ (or $\mathcal{N}=2^*$) $SU(2)$ SYM.
\vspace{15pt}

\subsection{Continuous q-Hermite polynomials \texorpdfstring{$(n_F=0)$.}{}}\label{app:qHermite}
It is well established that continuous q-Hermite polynomials play an important role in determining the spectrum of DSSYK \cite{Berkooz:2018qkz, Berkooz:2018jqr}. They are obtained from the q-Askey scheme by setting $t_1=t_2=t_3=t_4=0$. One can define them as:
\begin{flalign}
    H_n(\cos\theta|q^2) = \sum_{k=0}^n \frac{(q^2;q^2)_n}{(q^2;q^2)_k(q^2;q^2)_{n-k}}e^{i(n-2k)\theta}.
\end{flalign}
To achieve a chord diagram interpretation of the Schur half-indices, we heavily rely on the generating functions of q-Hermite polynomials \cite{Berkooz:2018qkz}:
\begin{flalign}
    &\sum_{n=0}^\infty H_n(\cos\theta|q^2) \frac{t^n}{(q^2;q^2)_n} = \frac{1}{(t e^{\pm i\theta};q^2)_\infty}\label{eqn:lin-qH-gf}
    \\&
    \sum_{n=0}^\infty H_n(\cos\theta_1|q^2)H_n(\cos\theta_2|q^2) \frac{t^n}{(q^2;q^2)_n} = \frac{(t^2;q^2)_\infty}{(t e^{\pm i(\theta+\phi)},t e^{\pm i(\theta-\phi)};q^2)_\infty}\label{eqn:bilin-qH-gf}
\end{flalign}
By appropriately rescaling q-Hermite polynomials, we build a set of orthonormal functions:
\begin{flalign}
    \psi_n(\cos\theta|q^2) \equiv \braket{n|\theta}_{n_F=0} = \sqrt{(q^2,e^{\pm 2i\theta};q^2)_\infty}\frac{H_n(\cos\theta|q^2)}{\sqrt{2\pi(q^2;q^2)_n}}\label{eqn:ort-qH}
\end{flalign}
Which satisfy the following $n$- and $\theta$-orthogonality relations:
\begin{flalign}
    &\int_0^\pi d\theta\; 
    \psi_n(\cos\theta | q^2) 
    \psi_m(\cos\theta | q^2 )=\delta_{n,m},
    \\&
    \sum_{n=0}^\infty 
    \psi_n(\cos\theta_1| q^2) 
    \psi_n(\cos\theta_2 | q^2) = \delta(\theta_1-\theta_2).
\end{flalign}
They obey the recursion relation $(\psi_n=\psi_n(\cos\theta|q^2))$:
\begin{equation}
    \frac{2\cos \theta}{\sqrt{1-q^2}} \psi_n = \sqrt{[n+1]_{q^2}} \psi_{n+1} + \sqrt{[n]_{q^2}} \psi_{n-1}.
\end{equation}
\vspace{15pt}

\subsection{Generalized q-Hermite polynomials (\texorpdfstring{$n_F=2$}{}).} 
\label{generalized q Hermite}

In this case, $(t_1,t_2,t_3,t_4)=(t_1,0,0,0)$. Generalized q-Hermite polynomials (also referred to as continuous big q-Hermite polynomials) can be expressed using the hypergeometric $_2\phi_0$ series as \cite{Koekoek2010}:
\begin{flalign}
    H_n(\cos\theta;t_1|q^2)=e^{in\theta} 
    \,_2\phi_0\bigg({q^{-2n}, t_1 e^{i\theta} \atop -}; q^2, q^{2n}e^{-2i\theta} \bigg),
\end{flalign}
and provide the simplest generalization of the ordinary q-Hermite polynomials. We now introduce a rescaled set of functions:
\begin{flalign}
    \phi_n(\cos\theta,t_1|q^2) \equiv \braket{n|\theta}_{n_F=2} = \sqrt{\frac{(q^2,e^{\pm 2i\theta};q^2)_\infty}{(t_1e^{\pm i\theta};q^2)_\infty}} \frac{H_n(\cos\theta;t_1|q^2)}{\sqrt{2\pi(q^2,q^2)_n}}.
\end{flalign}
Using known $n$- and $\theta$-orthogonality properties of  $H_n(\cos\theta;t_1|q^2)$ \cite{Koekoek2010}, one can show that polynomials $\phi_n$ satisfy:
\begin{flalign}
    &\int_0^\pi d\theta\; 
    \phi_n(\cos\theta; t_1 | q^2) 
    \phi_m(\cos\theta; t_1 | q^2 )=\delta_{n,m},
    \\&
    \sum_{n=0}^\infty 
    \phi_n(\cos\theta_1; t_1 | q^2) 
    \phi_n(\cos\theta_2; t_1 | q^2) = \delta(\theta_1-\theta_2).
\end{flalign}
The rescaled generalized q-Hermite polynomials $\phi_n=\phi_n(\cos\theta;t_1|q^2)$ satisfy:
\begin{equation}
    \frac{2\cos \theta}{\sqrt{1-q^2}} \phi_n = \sqrt{[n+1]_{q^2}} \phi_{n+1} + \frac{t_1 q^{2n}}{\sqrt{1-q^2}} \phi_n + \sqrt{[n]_{q^2}} \phi_{n-1}.
\end{equation}
\vspace{15pt}

\subsection{Al-Salam-Chihara polynomials (\texorpdfstring{$n_F=4$}{}).}
\label{app: al salam}

We now pick $(t_1,t_2,t_3,t_4) = (t_1,t_1^*,0,0)$. The Al-Salam-Chihara polynomials can be expressed using the $_3\phi_2$ hypergeometric function:
\begin{flalign}\label{eqn:ASC}
    Q_n(\cos\theta;t_1,t_2|q^2) = \frac{(t_1t_2;q^2)_n}{t_1^n} 
    \,_3 \phi_2\bigg({q^{-2n},t_1e^{\pm i\theta}\atop t_1t_2, 0}; q^2,q^2\bigg).
\end{flalign}
We define an orthonormal set of polynomials $\varphi_{\{n\}}$ as follows:
\begin{flalign}
    \varphi_n(\cos\theta;t_1,t_2|q^2) \equiv \braket{n|\theta}_{n_F=4} = \sqrt{\frac{(t_1t_2,q^2)_\infty}{(t_1t_2,q^2)_n}}
    \sqrt{\frac{(q^2,e^{\pm 2i\theta};q^2)_\infty}{(t_1e^{\pm i\theta}, t_2e^{\pm i\theta};q^2)_\infty}} \frac{Q_n(\cos\theta;t_1,t_2|q^2)}{\sqrt{2\pi (q^2;q^2)_n}}.
\end{flalign}
By using properties of Al-Salam-Chihara polynomials \cite{Koekoek2010,Berkooz:2018jqr}, one can show that $\varphi_n$ are orthogonal in $n$ and $\theta$:
\begin{flalign}
    &\int_0^\pi d\theta\; 
    \varphi_n(\cos\theta; t_1, t_2 | q^2)
    \varphi_m(\cos\theta; t_1, t_2 | q^2)=\delta_{n,m},
    \\&
    \sum_{n=0}^\infty 
    \varphi_n(\cos\theta_1; t_1, t_2 | q^2)
    \varphi_n(\cos\theta_2; t_1, t_2 | q^2) = \delta(\theta_1-\theta_2).
\end{flalign}
They satisfy the following recursion relation ($\varphi_n=\varphi_n(\cos\theta;t_1,t_2|q^2)$) \cite{Ismail2005}:
\begin{equation}\begin{aligned}\label{al Salam Chihara recursion}
    \frac{2\cos\theta}{\sqrt{1-q^2}} \varphi_n =& 
    \sqrt{[n+1]_{q^2}}\sqrt{1-t_1t_2 q^{2n}}\; \varphi_{n+1}
    + \sqrt{[n]_{q^2}}\sqrt{1-t_1t_2 q^{2n-2}}\; \varphi_{n-1} \\&
    + \bigg(\frac{t_1 q^{2n}}{\sqrt{1-q^2}} + \frac{t_2 q^{2n}}{\sqrt{1-q^2}}\bigg)\varphi_n.
\end{aligned}\end{equation}
\vspace{15pt}

\subsection{Continuous dual q-Hahn polynomials (\texorpdfstring{$n_F=6$}{}).} \label{app:qHahn}
We now pick $(t_1,t_2,t_3,t_4) = (t_1,t_1^*,t_3,0)$. One may define continuous dual q-Hahn polynomials using the $_3\phi_2$ hypergeometric series:
\begin{flalign}
    S_n(\cos\theta;t_1,t_2,t_3|q^2) = 
    \frac{(t_1t_2,t_1t_3;q^2)_n}{t_1^n} \,_3\phi_2\bigg({q^{-2n},t_1e^{\pm i\theta}\atop t_1t_2,t_1t_3};q^2,q^2\bigg).
\end{flalign}
It is convenient to rescale the continuous dual q-Hahn polynomials as follows:
\begin{flalign}\begin{aligned}
    &\chi_n(\cos\theta;t_1,t_2,t_3|q^2) \equiv \braket{n|\theta}_{n_F=6}= \\[5pt]&
    \sqrt{\frac{\prod_{1\leq i<j\leq 3} (t_it_j;q^2)_\infty}{\prod_{1\leq i<j\leq 3} (t_it_j;q^2)_n}}
    \sqrt{\frac{(q^2,e^{\pm 2i\theta};q^2)_\infty}{\prod_{1\leq i\leq 3}(t_i e^{\pm i\theta};q^2)_\infty}}\frac{S_n(\cos\theta;t_1,t_2,t_3|q^2)}{\sqrt{2\pi (q^2;q^2)_n}}.\label{eqn:qHahn}
\end{aligned}\end{flalign}

\noindent Their $\theta$-orthogonality relation can be derived from the asymmetric Poisson kernel $K_t^{t_i,\tau_i}(\theta_1,\theta_2)$ \cite{Askey:1996a} for continuous dual q-Hahn polynomials. We are not aware of any other place where this is shown, so we derive it here. We have:
\begin{flalign}
    \sum_{n=0}^\infty 
    (\tfrac{\tau_1}{t_1})^n \frac{t^n}{(\tau_1\tau_2,\tau_1\tau_3,t_2t_3,q^2;q^2)_n} S_n(\cos\theta_1;t_1,t_2,t_3|q^2)S_n(\cos\theta_2;\tau_1,\tau_2,\tau_3|q^2) = K_t^{t_i,\tau_i}(\theta_1,\theta_2).
\end{flalign}
A closed form expression for $K_t^{t_i,\tau_i}(\theta_1,\theta_2)$ is known in the literature \cite{Koelink:1999a,Askey:1996a} in terms of the very well poised $_8W_7$ series. We will only need the $t\to 1^-$ limit, in which the expression simplifies significantly:
\begin{flalign}\begin{aligned}
    &\sum_{n=0}^\infty (\tfrac{\tau_1}{t_1})^n
    \frac{1}{(\tau_1\tau_2,\tau_1\tau_3,t_2t_3,q^2;q^2)_n} S_n(\cos\theta_1;t_1,t_2,t_3|q^2) S_n(\cos\theta_2;\tau_1,\tau_2,\tau_3|q^2) =
    \\& \qquad \qquad \qquad \qquad \qquad \qquad  \qquad \qquad 
    \frac{(\tau_1 e^{\pm i\theta_1},\tau_2 e^{\pm i\theta_1},t_3 e^{\pm i\theta_2},(t_3/\tau_3)^2;q^2)_\infty}{(\tau_1\tau_2,t_1t_3,t_2t_3,\tfrac{t_3}{\tau_3}e^{\pm i\theta_1 \pm i \theta_2};q^2)_\infty}.
\end{aligned}\end{flalign}
Now by taking the limit $s_i=t_i/\tau_i\to 1$ and using the identity\footnote{The identity also contains a term proportional to $\delta(\theta_1+\theta_2)$, which we can safely neglect, since it gives a subleading contribution in $\int d\theta_1d\theta_2$.}:
\begin{flalign}\label{eqn:PKernel-limit}
    \lim_{s_3 \to 1} \frac{(s_3^2;q^2)_{\infty}}{(s_3e^{\pm i\theta_1 + \pm i\theta_2)};q^2)_{\infty}} = 2\pi \delta(\theta_1-\theta_2) \frac{1}{(e^{\pm 2i \theta_1};q^2)_{\infty} (q^2;q^2)_{\infty}},
\end{flalign}
we recover the $\theta$-orthogonality relation for the polynomials $S_n$:
\begin{flalign}\begin{aligned}\label{eqn:ort-S-t}
    &\sum_{n=0}^\infty \frac{1}{(t_1t_2,t_1t_3,t_2t_3,q^2;q^2)_n} S_n(\cos\theta_1;t_1,t_2,t_3| q^2) S_n(\cos\theta_2;t_1,t_2,t_3| q^2) = 
    \\& \qquad\qquad\qquad\qquad\qquad
    \frac{\prod_{i=1}^3 (t_ie^{\pm i\theta_1};q^2)_\infty}{\prod_{1\leq i<j\leq 3}(t_i t_j;q^2)_\infty(q^2,e^{\pm 2i\theta_1};q^2)_\infty} 2\pi\delta(\theta_1-\theta_2).
\end{aligned}\end{flalign}
The $n$-orthogonality relation is readily available in the literature \cite{Koekoek2010,Ismail2005,Gasper2004}: 
\begin{flalign}\begin{aligned}\label{eqn:ort-S-n}
    &\int_0^\pi d\theta \frac{(e^{\pm 2i\theta};q^2)_\infty}{\prod_{i=1}^3(t_ie^{\pm i\theta};q^2)_\infty}
    S_n(\cos\theta;t_1,t_2,t_3|q^2)
    S_m(\cos\theta;t_1,t_2,t_3|q^2) \;=\\&
    \qquad\qquad\qquad\qquad\qquad\qquad\qquad\qquad
    \frac{2\pi (q^2;q^2)_n \prod_{1\leq i<j\leq 3}(t_it_j;q^2)_n}{\prod_{1\leq i<j\leq 3}(t_it_j;q^2)_\infty (q^2;q^2)_\infty} \delta_{n,m}.
\end{aligned}\end{flalign}

\noindent Therefore, the two orthogonality relations for the rescaled continuous dual q-Hahn polynomials $\chi_n(\cos\theta;t_1,t_2,t_3|q^2)$ can be written as:
\begin{flalign}
    &\int_0^\pi d\theta\; 
    \chi_n(\cos\theta; t_1, t_2, t_3 | q^2)
    \chi_m(\cos\theta; t_1, t_2, t_3 | q^2)=\delta_{n,m},
    \\&
    \sum_{n=0}^\infty 
    \chi_n(\cos\theta_1; t_1, t_2, t_3 | q^2)
    \chi_n(\cos\theta_2; t_1, t_2, t_3 | q^2) = \delta(\theta_1-\theta_2).
\end{flalign}
The functions $\chi_n(\cos\theta;t_1,t_2,t_3|q^2)$ obey the following recursion relation:
\begin{equation}\begin{aligned}\label{eqn:qHahn-recursion}
    \frac{2\cos\theta}{\sqrt{1-q^2}} \chi_n =&
    \sqrt{[n+1]_{q^2}}\sqrt{\prod_{1\leq i<j\leq 3}(1-t_it_jq^{2n})} \chi_{n+1}\\&
    + \sqrt{[n]_{q^2}}\sqrt{\prod_{1\leq i<j\leq 3}(1-t_it_jq^{2n-2})} \chi_{n-1}\\&
    + \bigg(\frac{t_1+t_2+t_3}{\sqrt{1-q^2}}+\frac{t_1t_2t_3}{\sqrt{1-q^2}}(q^{2n-2}-q^{4n}-q^{4n-2})\bigg) \chi_n.
\end{aligned}\end{equation}
\vspace{15pt}

\subsection{Askey-Wilson polynomials (\texorpdfstring{$n_F=8$}{}).} 
Askey-Wilson polynomials are generally defined for $|t_i|\leq 1$, but for our purposes we restrict this quartet to complex conjugate pairs $(t_1,t_2,t_3,t_4)=(t_1,t_1^*,t_3,t_3^*)$. This family of polynomials is defined using the hypergeometric $_4\phi_3$ series:
\begin{flalign}\label{eqn:def-P}
    P_n(\cos\theta;t_1,t_2,t_3,t_4|q^2) = 
    \frac{(t_1t_2,t_1t_3,t_1t_4;q^2)_n}{t_1^n} \,_4\phi_3\bigg({q^{-2n},t_1t_2t_3t_4q^{2n-2},t_1e^{\pm i\theta}\atop t_1t_2,t_1t_3,t_1t_4};q^2,q^2\bigg)
\end{flalign}
We define a set of rescaled Askey-Wilson polynomials $\pi_n(\cos\theta; t_1, t_2, t_3, t_4 | q^2)$ as follows:
\begin{flalign}\begin{aligned}\label{eqn:ortn-AW}
    &\pi_n(\cos\theta;t_1,t_2,t_3,t_4|q^2) \equiv \; \braket{n|\theta}_{n_F=8} =\\[5pt]& 
    \sqrt{\frac{(1-t_1t_2t_3t_4 q^{4n-2})(t_1t_2t_3t_4q^{-2};q^2)_n}{(1-t_1t_2t_3t_4 q^{-2})\prod_{1\leq i<j\leq 4}(t_it_j;q^2)_n}}\\& \times
    \sqrt{\frac{\prod_{1\leq i<j\leq 4}(t_it_j;q^2)_\infty (q^2,e^{\pm 2i\theta};q^2)_\infty}{(t_1t_2t_3t_4;q^2)_\infty \prod_{i=1}^4 (t_ie^{\pm i \theta};q^2)_\infty}}\frac{P_n(\cos\theta;t_1,t_2,t_3,t_4|q^2)}{\sqrt{2\pi(q^2;q^2)_n}}.
\end{aligned}\end{flalign}
As we now demonstrate, $\pi_n$ satisfy $\theta$- and $n$-orthogonality conditions with unit measure.\\

\noindent The $\theta$-orthogonality relation can be derived starting from the asymmetric Kernel for Askey-Wilson polynomials:
\begin{flalign}\begin{aligned}
    &\sum_{n=0}^\infty \frac{(1-t_1t_2t_3t_4q^{4n-2})}{(1-t_1t_2t_3t_4q^{-2})}\frac{(t_1t_2t_3t_4q^{-2};q^2)_n \tau_1^n t_1^{-n} t^n}{(q^2;q^2)_n\prod_{1\leq i<j\leq 4}(t_it_j;q^2)_n} P_n(\cos\theta_1;t_i|q^2)P_n(\cos\theta_2;\tau_i|q^2)\\[5pt]& = K_t^{t_i,\tau_i}(\theta_1,\theta_2|q^2).
\end{aligned}\end{flalign}
The Kernel $K_t^{t_i,\tau_i}(\theta_1,\theta_2|q^2)$ is expressed in terms of multiple very-well-poised $_8W_7$ hypergeometric series \cite{Askey:1996a}. However, in the $t\to 1^-$ limit, the expression simplifies to:
\begin{flalign}\begin{aligned}
    &\sum_{n=0}^\infty \frac{(1-t_1t_2t_3t_4q^{4n-2})}{(1-t_1t_2t_3t_4q^{-2})}\frac{(t_1t_2t_3t_4q^{-2};q^2)_n}{(q^2;q^2)_n\prod_{1\leq i<j\leq 4}(t_it_j;q^2)_n} P_n(\cos\theta_1;t_i|q^2)P_n(\cos\theta_2;\tau_i|q^2) 
    \;=\\[5pt]&
    \frac{(t_1t_2t_3t_4;q^2)_\infty}{(\tau_1\tau_2,t_1t_3,t_1t_4,t_2t_3,t_2t_4,t_3t_4;q^2)_\infty} 
    \frac{(\tau_1e^{\pm i\theta_1},\tau_2e^{\pm i\theta_1},\tau_3e^{\pm i\theta_2},\tau_4e^{\pm i\theta_2},(\tfrac{\tau_2}{t_2})^2;q^2)_\infty}{(\tfrac{\tau_2}{t_2}e^{\pm i\theta_1\pm i \theta_2};q^2)_\infty}.
\end{aligned}\end{flalign}
\vspace{2pt}

\noindent One can now take the $t_i/\tau_i\to 1$ limit. Using \eqref{eqn:PKernel-limit}, we find that the kernel localizes to a $\delta$-function:
\begin{flalign}\begin{aligned}\label{eqn:ort-P-t}
    &\sum_{n=0}^\infty \frac{(1-t_1t_2t_3t_4q^{4n-2})}{(1-t_1t_2t_3t_4q^{-2})}\frac{(t_1t_2t_3t_4q^{-2};q^2)_n}{(q^2;q^2)_n\prod_{1\leq i<j\leq 4}(t_it_j;q^2)_n} P_n(\cos\theta_1;t_i|q^2)P_n(\cos\theta_2;t_i|q^2) 
    \;=\\[5pt]&
    \frac{(t_1t_2t_3t_4;q^2)_\infty \prod_{i=1}^4(t_ie^{\pm i\theta_1};q^2)_\infty}{\prod_{1\leq i<j\leq 4}(t_it_j;q^2)_\infty (q^2,e^{\pm 2i\theta_1};q^2)_\infty} 2\pi\delta(\theta_1-\theta_2) .
\end{aligned}\end{flalign}
\noindent The $n$-orthogonality relation is known in literature \cite{Askey:1996a,Koekoek2010,Ismail2005} and for our choice of parameters $t_i$ reduces to:
\begin{flalign}\begin{aligned}\label{eqn:ort-P-n}
    &\int_0^\pi d\theta \frac{(e^{\pm 2i\theta};q^2)_\infty}{\prod_{i=1}^4(t_ie^{\pm i\theta};q^2)_\infty}\frac{1}{(t_1t_2,t_1t_3,t_1t_4;q^2)_m} P_n(\cos\theta;t_i|q^2)P_m(\cos\theta;t_i|q^2) \;=\\&
    \frac{(1-t_1t_2t_3t_4q^{-2})(q^2,t_3t_4,t_2t_4,t_2t_3;q^2)_n}{(1-t_1t_2t_3t_4 q^{4n-2}) (t_1t_2t_3t_4q^{-2};q^2)_n} \frac{2\pi(t_1t_2t_3t_4;q^2)_\infty}{\prod_{1\leq i<j\leq 4}(t_it_j;q^2)_\infty (q^2;q^2)_\infty} \delta_{n,m}.
\end{aligned}\end{flalign}

\vspace{5pt}
\noindent Using \eqref{eqn:ortn-AW}, one can now show that the orthogonality relations \eqref{eqn:ort-P-n}, \eqref{eqn:ort-P-t} become:
\begin{flalign}
    &\int_0^\pi d\theta\; \pi_n(\cos\theta; t_1, t_2, t_3, t_4 | q^2)\pi_m(\cos\theta; t_1, t_2, t_3, t_4 | q^2)=\delta_{n,m},
    \\&
    \sum_{n=0}^\infty 
    \pi_n(\cos\theta_1; t_1, t_2, t_3, t_4 | q^2)
    \pi_n(\cos\theta_2; t_1, t_2, t_3, t_4 | q^2) = \delta(\theta_1-\theta_2).
\end{flalign}
$\pi_n(\cos\theta; t_1, t_2, t_3, t_4 | q^2)$ satisfy a tedious three-term recursion relation \cite{Koekoek2010}. We write it out, keeping the $t_{1,2,3,4}$ dependence explicit:
\begin{flalign}\begin{aligned}\label{eqn:AW_tmatrix}
    &\frac{2\cos\theta}{\sqrt{1-q^2}}\pi_n = \\& 
    = \bigg(\sqrt{[n+1]_{q^2}}\frac{1}{(1-t_1t_2t_3t_4q^{4n})}\sqrt{\frac{\prod_{1\leq i<j\leq 4}(1-t_it_jq^{2n})(1-t_1t_2t_3t_4q^{2n-2})}{(1-t_1t_2t_3t_4q^{4n-2})(1-t_1t_2t_3t_4q^{4n+2})}}\bigg)\pi_{n+1} 
    \;\\[15pt]& 
    +\bigg(\sqrt{[n]_{q^2}}\frac{1}{(1-t_1t_2t_3t_4q^{4n-4})}\sqrt{\frac{\prod_{1\leq i<j\leq 4}(1-t_it_jq^{2n-2})(1-t_1t_2t_3t_4q^{2n-4})}{(1-t_1t_2t_3t_4q^{4n-6})(1-t_1t_2t_3t_4q^{4n-2}) }}\bigg) \pi_{n-1}
    \\[15pt]& + 
    \bigg(t_1+t_1^{-1}-\frac{(1-t_1t_2q^{2n})(1-t_1t_3q^{2n})(1-t_1t_4q^{2n})(1-t_1t_2t_3t_4 q^{2n-2})}{t_1(1-t_1t_2t_3t_4q^{4n-2})(1-t_1t_2t_3t_4q^{4n})}\\& \qquad\qquad
    -\frac{t_1(1-q^{2n})(1-t_2t_3q^{2n-2})(1-t_2t_4q^{2n-2})(1-t_3t_4q^{2n-2})}{(1-t_1t_2t_3t_4q^{4n-4})(1-t_1t_2t_3t_4q^{4n-2})}\bigg)\frac{1}{\sqrt{1-q^2}}\pi_n.
\end{aligned}\end{flalign}
\vspace{15pt}

\subsection{Continuous q-ultraspherical polynomials (adjoint, \texorpdfstring{$\mathcal{N}=2^*$}{}).} \label{sec:qUltraspherical}
Also referred to as Rogers polynomials, this family of polynomials is obtained by setting $(t_1,t_2,t_3,t_4)=(\sqrt{\beta},\sqrt{\beta} q,-\sqrt{\beta},-\sqrt{\beta} q)$. One can define them via the $_4\phi_3$ hypergeometric series:
\begin{flalign}
    C_n(\cos\theta;\beta|q^2) = 
    \frac{(\beta^2 ;q^2)_n}{(q^2;q^2)_n} \beta^{-\frac{n}{2}}\,_4\phi_3\bigg({q^{-2n},\beta^2q^{2n},\beta^{1/2}e^{\pm i\theta} \atop \beta q,-\beta,-\beta q};q^2,q^2\bigg).
\end{flalign}
The continuous q-ultraspherical polynomials are directly related to Askey-Wilson polynomials \eqref{eqn:def-P} as:
\begin{flalign}
    C_n(\cos\theta;\beta | q^2) = \frac{(\beta^2;q^2)_n}{(\beta q,-\beta,-\beta q,q^2;q^2)_n}P_n(\cos\theta; \sqrt{\beta},\sqrt{\beta}q,-\sqrt{\beta},-\sqrt{\beta}q | q^2).
\end{flalign}
They therefore satisfy analogous orthogonality relations to \eqref{eqn:ort-P-t}, \eqref{eqn:ort-P-n}. We find:
\begin{flalign}\begin{aligned}\label{eqn:ort-C-t}
    &\sum_{n=0}^\infty \frac{1-\beta^2q^{4n}}{1-\beta^2}\frac{(-\beta,q^2;q^2)_n}{(-\beta q^2,\beta^2;q^2)_n} C_n(\cos\theta_1;\beta | q^2)C_n(\cos\theta_2;\beta | q^2) =\\&
    \frac{(\beta^2 q^2;q^2)_\infty}{(-\beta,-\beta q^2;q^2)_\infty (\beta^2 q^2;q^4)_\infty^2}
    \frac{(\beta e^{\pm 2i\theta_1},\beta q^2 e^{\pm 2i\theta_1};q^4)_\infty}{(q^2,e^{\pm 2i\theta_1})_\infty} 2\pi\delta(\theta_1-\theta_2),
\end{aligned}\end{flalign}
\vspace{7pt}
\begin{flalign}\begin{aligned}\label{eqn:ort-C-n}
    &\int_0^\pi d\theta \frac{(e^{\pm 2i\theta};q^2)_\infty}{(\beta e^{\pm 2i\theta},\beta q^2 e^{\pm 2i\theta_1};q^4)\infty} \frac{(-\beta,q^2;q^2)_m}{(\beta^2;q^2)_m} C_n(\cos\theta;\beta | q^2)C_m(\cos\theta;\beta | q^2) =\\&
    \qquad\qquad\qquad\qquad\qquad
    \frac{1-\beta^2}{1-\beta^2 q^{4n}} \frac{2\pi (\beta^2 q^2;q^2)_\infty}{(q^2;q^2)_\infty (\beta^2 q^2;q^4)_\infty^2(-\beta,-\beta q^2;q^2)_\infty} \delta_{n,m}.
\end{aligned}\end{flalign}
We now define the rescaled polynomials:
\begin{flalign}\begin{aligned}
    &\zeta_n(\cos\theta;\beta|q^2) \equiv \braket{n|\theta}_{n_F=2, adj.} = \\[5pt]&
    \qquad\qquad
    \sqrt{\frac{1-\beta q^{2n}}{1-\beta}
    \frac{(q^2;q^2)_n}{(\beta^2;q^2)_n}
    \frac{(\beta^2 q^2;q^4)^2_\infty (-\beta,-\beta q^2;q^2)_\infty}{(\beta^2 q^2;q^2)_\infty}}\\& \qquad\qquad \times
    \sqrt{\frac{(q^2 e^{\pm 2i\theta};q^2)_\infty}{2\pi(\beta e^{\pm 2i\theta},\beta q^2 e^{\pm 2i\theta};q^4)_\infty}} 
    C_n(\cos\theta;\beta | q^2)\label{eqn:zeta_n}
\end{aligned}\end{flalign}
From \eqref{eqn:ort-C-t}, \eqref{eqn:ort-C-n}, we find that $\zeta_n$ satisfy following $n$- and $\theta$-orthogonality relations:
\begin{flalign}
    &\int_0^\pi d\theta\; 
    \zeta_n(\cos\theta; \beta | q^2)
    \zeta_m(\cos\theta; \beta | q^2)=\delta_{n,m},\label{eqn:ort-zeta-n}
    \\&
    \sum_{n=0}^\infty 
    \zeta_n(\cos\theta_1; \beta | q^2)
    \zeta_n(\cos\theta_2; \beta | q^2) = \delta(\theta_1-\theta_2).\label{eqn:ort-zeta-t}
\end{flalign}
Polynomials $\zeta_n(\cos\theta; \beta | q^2)$ satisfy a recursion relation, without a diagonal term \cite{Koekoek2010}:
\begin{flalign}\begin{aligned}
    \frac{2\cos\theta}{\sqrt{1-q^2}}\zeta_n =& 
    \sqrt{[n+1]_{q^2}}\sqrt{\frac{1-\beta^2 q^{2n}}{(1-\beta q^{2n})(1-\beta q^{2n+2})}}\zeta_{n+1}
    \\& 
    + \sqrt{[n]_{q^2}}\sqrt{\frac{1-\beta^2 q^{2n-2}}{(1-\beta q^{2n-2})(1-\beta q^{2n})}}\zeta_{n-1}\label{eqn:zetaRR}
\end{aligned}\end{flalign}

\bibliographystyle{JHEP}
\bibliography{biblio.bib}

\end{document}